\documentclass[lineno]{article}
\usepackage{arxiv}

\usepackage{graphicx}
\usepackage{newtxtext}
\usepackage{newtxmath}
\usepackage{natbib}
\usepackage{hyperref}
\hypersetup{
    colorlinks = true,
    urlcolor   = blue,
    citecolor  = black,
}

\usepackage{graphicx}
\usepackage{epstopdf, epsfig}
\usepackage{color}
\usepackage{bm}
\usepackage[T2A]{fontenc}			
\usepackage[utf8]{inputenc}
\usepackage[english]{babel}

\newcommand\Ru{\mbox{\textit{Re}}} 

\newcommand\Ha{\mbox{\textit{Ha}}} 
\newcommand\N{\mbox{\textit{N}}} 
\newcommand\uxa{\langle u_x \rangle}

\begin{document}


\title{Experimental study of submerged liquid metal jet in transverse magnetic field}

\author{ \href{https://orcid.org/0000-0002-6397-3251}{\includegraphics[scale=0.06]{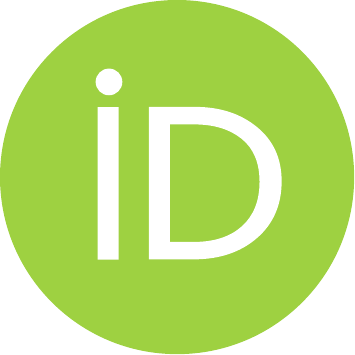}\hspace{1mm}Ivan A.~Belyaev}\\
		Joint Institute for High Temperature RAS,\\ Izhorskaya 13 Bd.2, 125412, Moscow,  Russia\\
	\texttt{bia@ihed.ras.ru} \\
	\And
	{Ivan  S.~Mironov} \\
	Joint Institute for High Temperature RAS,\\ Izhorskaya 13 Bd.2, 125412, Moscow,  Russia\\
	\And
	{Nikita  A.~Luchinkin} \\
	Joint Institute for High Temperature RAS,\\ Izhorskaya 13 Bd.2, 125412, Moscow,  Russia\\
	\And
	\href{https://orcid.org/0000-0002-6288-9787}{\includegraphics[scale=0.06]{orcid.pdf}\hspace{1mm}Yaroslav  I.~Listratov} \\
	Moscow Power Engineering Institute,\\ 
	Krasnokazarmennaya 14, 111250, Moscow,  Russia\\
	\And
	\href{https://orcid.org/0000-0003-1840-0232}{\includegraphics[scale=0.06]{orcid.pdf}\hspace{1mm}Yuri  B.~Kolesnikov} \\
	Technische Universitat Ilmenau,\\
	PF 100565, 98684 Ilmenau,Germany\\
	\And
	{Dmitry Kransnov} \\
	Technische Universitat Ilmenau,\\
	PF 100565, 98684 Ilmenau,Germany\\
	\And
	\href{https://orcid.org/0000-0003-3844-1779}{\includegraphics[scale=0.06]{orcid.pdf}\hspace{1mm}Oleg Zikanov} \\
	University of Michigan Dearborn,\\
	Dearborn, MI 48128-1491, USA\\
	\and
	{Sergey Molokov} \\
	Technische Universitat Ilmenau,\\
	PF 100565, 98684 Ilmenau,Germany\\
	}



\date{November 05, 2021}
\maketitle

\begin{abstract}
A liquid metal flow in the form of a submerged round jet entering a square duct in the presence of a transverse magnetic field is studied experimentally. A range of high Reynolds and Hartmann numbers is considered. Flow velocity is measured using electric potential difference probes. A detailed study of the flow in the duct's cross-section about seven jet's diameters downstream of the inlet reveals the dynamics, which is unsteady and dominated by high-amplitude fluctuations resulting from instability of the jet. The flow structure and fluctuation properties are largely determined by the value of the Stuart number $\N$. At moderate $\N$, the mean velocity profile retains a central jet with three-dimensional perturbations increasingly suppressed by the magnetic field as $\N$ grows. At higher values of $\N$, the flow becomes quasi-two-dimensional and acquires the form of an asymmetric macrovortex, with high-amplitude velocity fluctuations reemerging.
\end{abstract}


\section{Introduction}
Fusion is considered to be a future source of clean energy. It does not involve nuclear waste characteristic for fission. The fuel for fusion, a combination of the isotopes of hydrogen deuterium and tritium, are either abundant in the sea water or can be generated (bred) as a byproduct of the reaction. In the reactor, these isotopes are heated up to the temperature of several million Kelvin until the fusion reaction starts. The burning plasma in the reactor is contained by a very high magnetic field of 5-10T. The plasma chamber is surrounded by the blankets, which may employ liquid metals, Li or  PbLi, and which act simultaneously as heat exchangers diverting reaction's energy into an external circuit, shields, and breeders of tritium \citep{buhler2007liquid,Abdou:2015}. Since liquid metals are electrically conducting, their flow is strongly affected by the magnetohydrodynamic (MHD) interaction.

\begin{figure}
\includegraphics[width=0.99\textwidth]{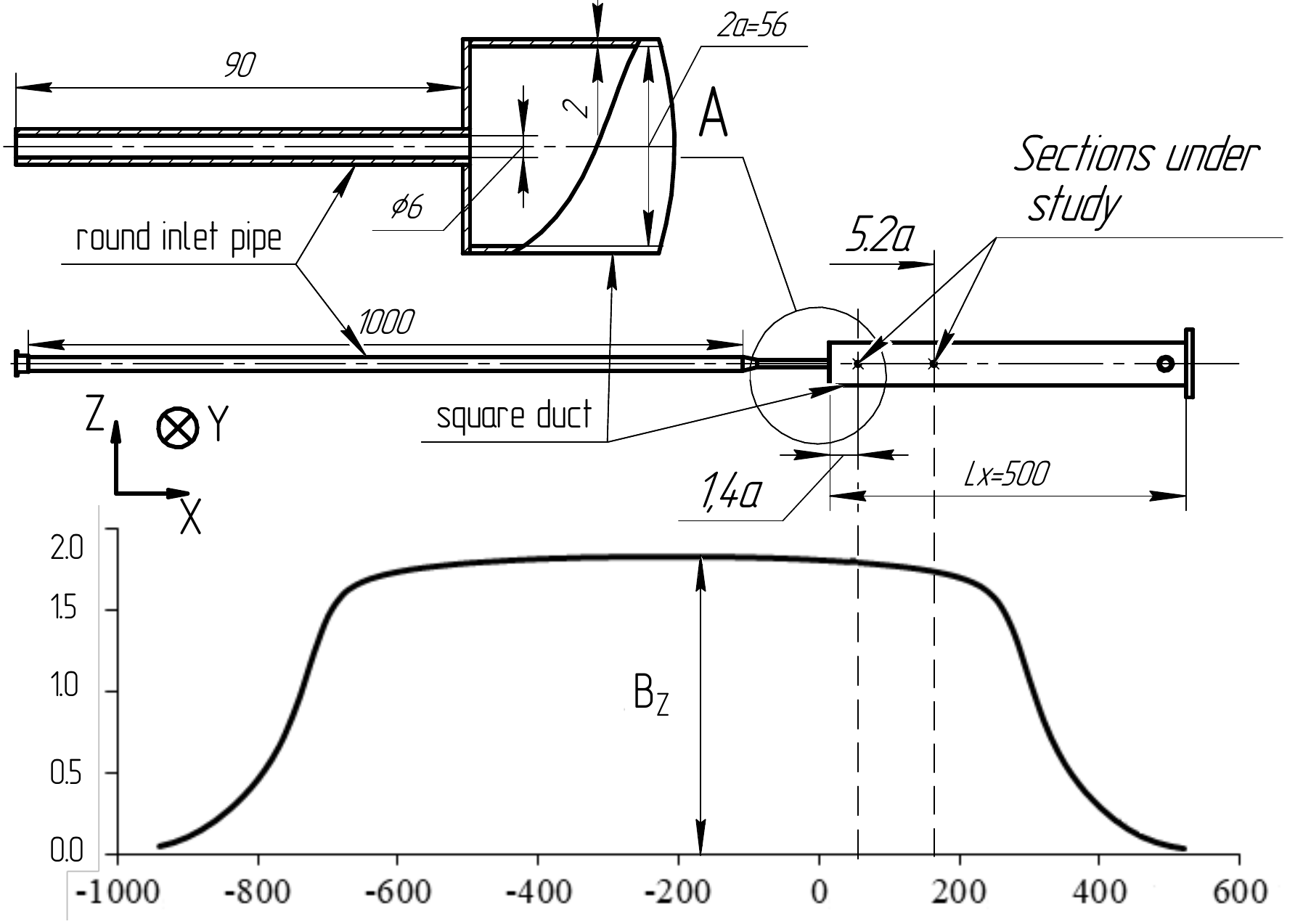}
\caption{Configuration of the experiment: the test section and the $x$-distribution of the main, $B_z$, component of the imposed magnetic field. The distances are given in mm, while the magnetic field - in Tesla. An inlet pipe of length 1 m is necessary to position the experimental section in the magnet.} 
\label{fig:DuctConfiguration}
\end{figure}

Liquid metal blankets contain many elements, such as straight ducts, bends, manifolds, etc. One of the key elements of the blanket is an expansion of a circular pipe into a larger rectangular duct (see figure \ref{fig:DuctConfiguration}). Understanding the flow dynamics in the expansion is important because it largely contributes to the flow distribution in a subsequent manifold.  If both the ratio of the duct to pipe dimensions and the fluid velocity are high, the resulting flow is in effect an initially circular jet entering a rectangular duct under the influence of a high, transverse magnetic field. 

Another important technology, in which a similar flow structure is observed is the electromagnetic braking in continuous casting of steel. Molten steel entering the solidification mold through a nozzle forms a submerged jet in a rectangular domain. As explained, for example by \citet{cukierski2008flow}, a strong transverse magnetic field is often applied to suppress excessive turbulent fluctuations around the jet and prevent strong impingement of the jet into the wall.  

The flow is also interesting from the fundamental point of view. It can be classified as one of several archetypal configurations of liquid metal MHD, other such configurations being, for example, a duct flow in a uniform or non-uniform transverse magnetic field, an internal shear layer, an MHD boundary layer, or magnetoconvection in an enclosure or a duct. As we will see in this paper, the jet flow demonstrates and helps to understand fundamental properties of flow transformation caused by an imposed magnetic field. In particular, the flow remains turbulent even in a high magnetic field as it evolves from three- to quasi-two-dimensional form.

Instability and transition to turbulence in axisymmetric and planar jet of viscous fluids without the magnetic field effect is a thoroughly studied area of fluid mechanics (see, e.g., \citet{michalke1984survey,thomas1986structural,stanley2002study}). Jets are known to be unstable at all, but smallest values of the Reynolds number \citep{sato1960stability,sato1964experimental}. The instability mechanism is that of the Kelvin-Helmholtz type and leads to vortices in  strongly sheared outer regions of a jet and jet's rapid distortion and development of turbulence. The process is strongly affected by inlet conditions (shape of the inlet, velocity profile, and presence of perturbations).

A sufficiently strong imposed magnetic field completely changes jet flow behaviour if fluid is electrically conducting. An example of this is the experiments with a submerged annular jet in an axial uniform magnetic field \citep{kalis1984stability,Klyukin1993}. Here, the transition to turbulence also occurs through the Kelvin-Helmholtz type instability, but the process demonstrates unique MHD features, such as development of a quasi-two-dimensional state and pronounced hysteresis in the plane of the Reynolds and Hartmann numbers \citep{kljukinKolesnikov1989}.

Flows of laminar, inviscid liquid metal jets in an unbounded fluid in a transverse, uniform magnetic field $\mathbf{B}$ were theoretically studied by \citet{moffatt1967annihilation}  and \citet{davidson1995magnetic}. MHD effects in these flows can be understood by considering the electrodynamic part of the MHD equations \citep{muller2001magnetofluiddynamics}, namely Ohm's law

\begin{equation}\label{Ohms}
\mathbf{j}=\sigma (-\nabla \varphi +\mathbf{u}\times \mathbf{B}),
\end{equation}
and the charge conservation equation

\begin{equation}
\nabla \cdot \mathbf{j=0}.
\end{equation}
Here  $\mathbf{u}=(u_{x},u_{y},u_{z})$ is the fluid velocity,  $\mathbf{j}=(j_{x},j_{y},j_{z})$ is the electric current density, $\varphi $ is the electric potential, $\sigma $ is the electrical conductivity of the fluid, and $(x,y,z)$ are the Cartesian co-ordinates. The Lorentz force acting on the fluid is: 
\begin{equation}
\mathbf{f}=\mathbf{j}\times \mathbf{B}.
\end{equation}

\citet{moffatt1967annihilation} considered a two-dimensional ($y$-independent) jet flowing in the $x-$direction in a transverse magnetic field $\mathbf{B}=B\mathbf{e}_{z}$. In a two-dimensional flow $\varphi =$ constant, so that the first term in the brackets in (\ref{Ohms}) vanishes. Then $\mathbf{j}=\sigma \mathbf{u}\times \mathbf{B}=-\sigma u_{x}B\mathbf{e}_{y}$, the Lorentz force, $\mathbf{f}=-\sigma u_{x}B^{2}\mathbf{e}_{x}$, reduces to pure braking at every point within the flow, and the axial component of momentum is annihilated by the field.

\begin{figure}
\begin{center}

\includegraphics[width=0.7\textwidth]{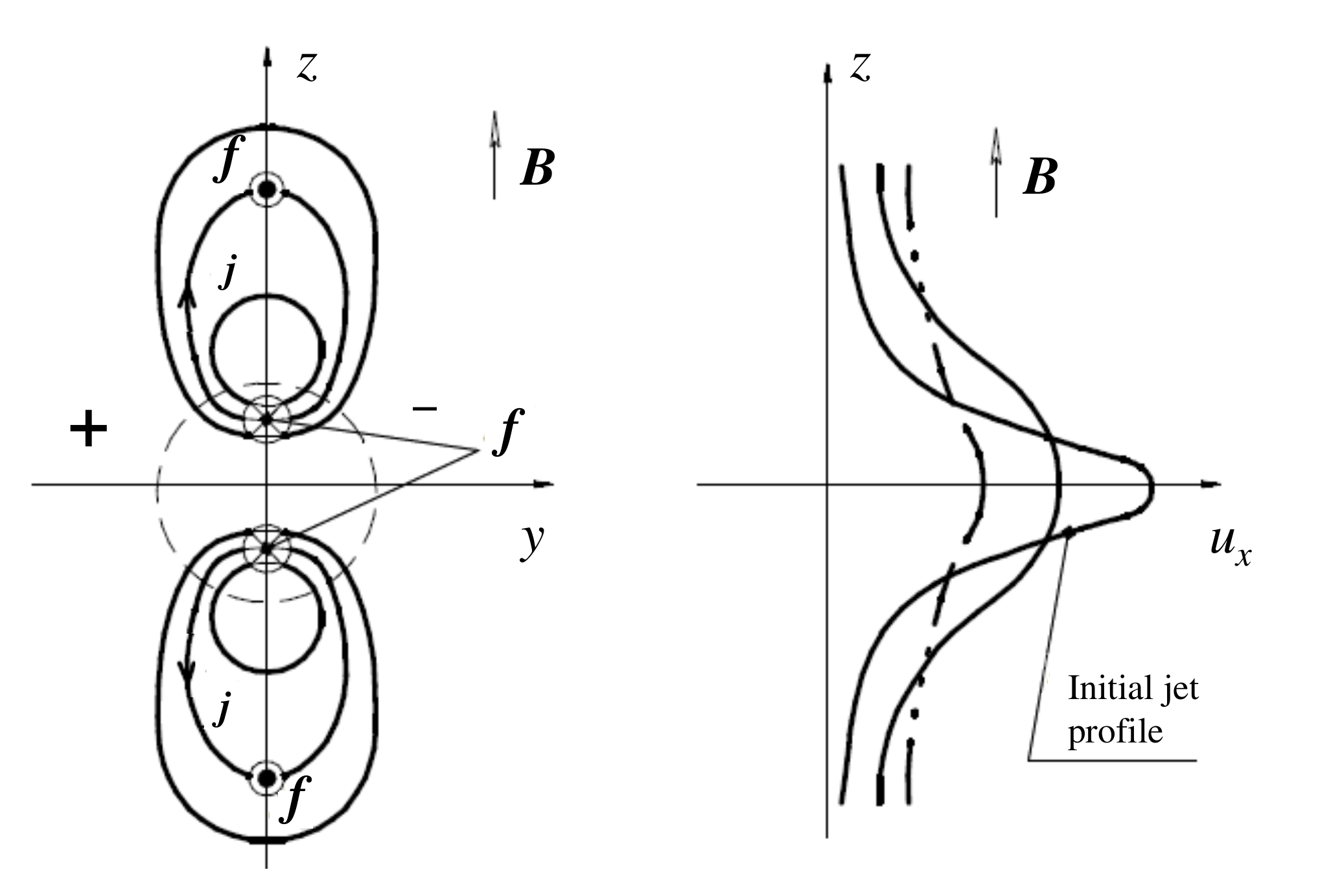}
\begin{minipage}[h]{1\linewidth}
\begin{tabular}{p{0.3\linewidth}p{0.25\linewidth}p{0.30\linewidth}}
\raggedleft \it{(a)} &  & \raggedright \it{(b)}\\
\end{tabular}
\end{minipage}
 \caption{Schematic diagram of the jet expansion along the magnetic field lines: an initially round infinite jet (broken line), electric current lines, and the directions of the Lorentz force (a), evolution of the jet profiles in the centre-plane $y=0$ as time increases (b).} 
\label{fig:JetCurandFor}
\end{center}
\end{figure}

If the jet is not two-dimensional, axial momentum is conserved in the zero-viscosity limit, but is redistributed along the field lines by the Lorentz force \citep{davidson1995magnetic} as shown in figure \ref{fig:JetCurandFor}. To understand the process we consider a straight, infinite, initially circular jet flowing in the $x-$direction. Now $\varphi $ is a function of both $y$ and $z$. In the original jet cross-section, currents are induced by the electromotive force $\mathbf{u}\times \mathbf{B}$, and the resulting Lorentz force brakes the flow. The currents induce the electric potential difference between the left and right parts of the jet shown by $'+'$ and $'-'$ in figure \ref{fig:JetCurandFor}a. This difference in the electric potential drives the current back through the fluid from $'+'$ to $'-'$ as shown in the figure. The Lorentz force accelerates the fluid above and below the original jet. The process continues with time as more and more layers of fluid become involved in the motion. The development of the axial velocity profiles with time are shown schematically in figure \ref{fig:JetCurandFor}b. The resulting jet has an anisotropic structure. We will show below that this evolution is very important for understanding of the flow pattern. It should be noted that similar anisotropic structures are found in many MHD flows, including flows around solid bodies \citep[see, e.g.,][]{ludford1960inviscid,hasimoto1960steady}, rotation of vortices \citep{davidson1995magnetic} and solid bodies \citep{molokov1993single}, electrically driven flows \citep{moffatt1964electrically}, etc. 

It is evident that the just outlined scenario of jet transformation does not form a complete picture. Inertial effects, hydrodynamic instabilities, and turbulence anticipated at sufficiently high Reynolds numbers are not included. Not less importantly, the effect of walls surrounding the flow domain and altering paths of electric currents is expected to be strong, much more so than in a flow without magnetic field. The effect remains poorly understood. 

Surprisingly, very little work has been done to analyze the flow. To our best knowledge, no experiments have been performed in the last three decades. The only computational work addressing the conclusions of \citet{davidson1995magnetic} concerning the transformation of a round jet by a transverse magnetic field was the large eddy simulations of  \citet{kim2004large}. The closest we can find to a study of MHD jet in a wall-bounded domain is the recent work of \citet{krasnov2021transformation} and \citet{belyaevflat}, in which transformation of a planar jet in a transverse magnetic field is analyzed.

The aim of this study is to understand the dynamics of a round jet as it enters a square duct. The study is divided into two parts. Here, in Part 1, we present the experimental results. Theoretical results using Direct Numerical Simulation will be presented in Part 2 by \cite{krasnov2022}. 

\section{MHD transformation of a submerged jet in a duct}
MHD flows of liquid metals in ducts are governed by four dimensionless parameters: the Reynolds number, $\Ru$, the Hartmann number, $Ha$, the Stuart number, $N=Ha^{2}/\Ru$, also known as the interaction parameter, and the wall conductance ratio, $c$. They characterise the ratios of inertial to viscous, electromagnetic to viscous, electromagnetic to inertial forces, and the ratio of wall to fluid electrical conductances, respectively. Their definitions are given in Table \ref{tab:RK-3}.

\begin{figure}
\includegraphics[width=1\textwidth]{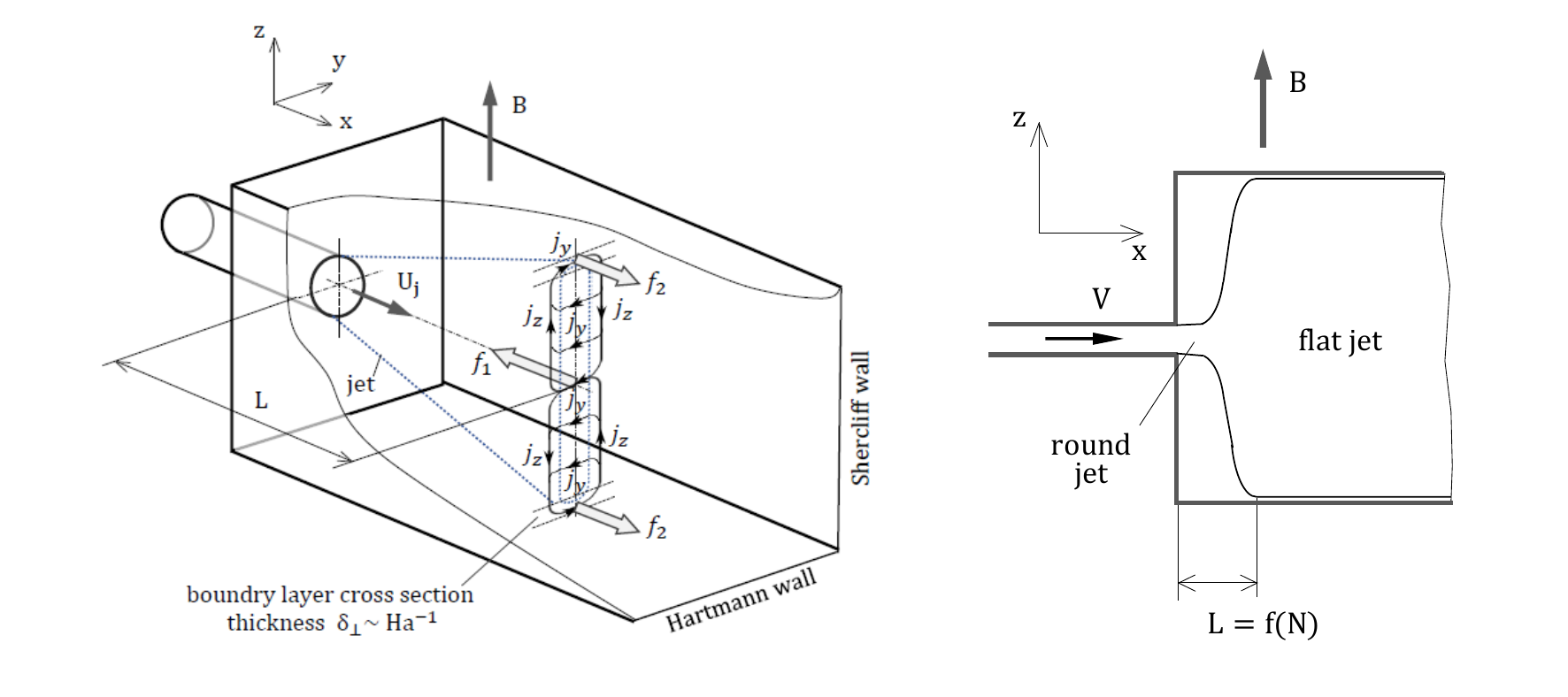}
\begin{minipage}[h]{1\linewidth}
\begin{tabular}{p{0.30\linewidth}p{0.35\linewidth}p{0.30\linewidth}}
\raggedleft \it{(a)} &  & \raggedright \it{(b)}\\
\end{tabular}
\end{minipage}
\caption{Two-dimensionalization of a round jet by a transverse magnetic field in a duct.}
\label{fig:JetAnnihilation}
\end{figure}

Consider a flow in a duct (figure \ref{fig:JetAnnihilation}a). The pairs of walls transverse and parallel to the magnetic field are called Hartmann walls and the sidewalls, respectively. In a sufficiently strong magnetic field  Hartmann layers of non-dimensional thickness $\sim Ha^{-1}$ are formed at the Hartmann walls. The streamwise velocity has an exponential profile inside these layers as determined by the balance between the viscous and Lorentz forces. There are also Shercliff (sidewall) layers of thickness $\sim Ha^{-1/2}$ forming at the sidewalls, but they seem to play no role in this study.

Consider now a flow created specifically by a round jet entering duct. From the discussion in the previous section it is clear that it will be transformed by the action of the electromagnetic forces. The transformation is in space, with the flow evolving with the downstream distance. Initially, the electric current lines are similar to those shown in figure \ref{fig:JetCurandFor}a. As the jet progresses through the duct, the gradients of the velocity profile along the magnetic field become smaller, and the jet develops anisotropy. Further downstream, the gradients virtually disappear. The velocity profile becomes almost independent of the $z-$coordinate except for thin Hartmann layers. The jet becomes quasi-two-dimensional, planar-like at a distance $L$ from the inlet as illustrated in figure \ref{fig:JetAnnihilation}. If Hartmann walls are electrically insulating, all the lines of the electric current will pass through these layers. They are very thin, and their electrical resistance is inversely proportional to their thickness, i.e. scales as $Ha$. Because of the increased resistance, the induced currents become much smaller in magnitude, and the resulting Hartmann damping decreases. If Hartmann walls are electrically conducting, part of the current will flow in the walls as well. Thus its magnitude increases, and damping increases as a result. 

In a three-dimensional flow, the longitudinal component of electric current $j_x$ also appears due to the variation of the electric potential along the flow \citep{muller2001magnetofluiddynamics}. Indeed, in the pipe the average fluid velocity is higher than in the duct. Thus, $\mathbf{u}\times \mathbf{B}$ is higher as well, and the induced potential difference in the $(x,y)-$planes is higher. These currents result in the Lorentz force stretching the jet in the $(x,y)-$planes towards the sidewalls, but the importance of this effect in this experimental study is unclear.

\section{Description of experiment}
\label{sec:description}
Experiments are performed using the HELME facility (HELMEF) described by \citet{Belyaev2017}.  Mercury  circulating in a closed loop is used as a model liquid. Walls are made of stainless steel and have thickness of 2 mm. The test-section is placed inside of an electromagnet, which can create the magnetic field of up to $1.8$ Tesla.  

The test section is configured as illustrated in figure \ref{fig:DuctConfiguration}. Mercury enters through an inlet into a square duct of half-height $a=28$ mm and length $L_x=500$ mm. The inlet has the form of a round pipe, 90 mm long with 6 mm inner diameter located at the mid-plane of the duct (figure \ref{fig:DuctConfiguration}a). The flow is driven in the $x$-direction by a pump. An imposed magnetic field $\bm{B}$ has the dominant component along the  $z$-direction. The uniform part of the magnetic field covers approximately the first $ 250$ mm of the duct length. 

In order to avoid possible disturbances caused by gas bubbles in the test section, the loop is vacuumed prior to filling it with mercury. Furthermore, experiments are repeated, with consistent results, with the test section oriented both vertically and horizontally.

A range of flow rates from 0.75 l/min to 5.4 l/min corresponding to the duct Reynolds numbers $600\le \Ru \le 7000$ is covered in the experiments. The bottom limit is set by the impossibility of accurate measurements of flow velocity occurring at lower flow rates due to the significant impact of electromagnetic noise. The top limit corresponds to the beginning of cavitation in the jet area.
The mean flow velocity $U_0$ used to define the duct Reynolds number is based on the flow rate measured using an electromagnetic flow meter. Flow stabilizer with gas cavity is used to avoid possible flow fluctuations produced by the pump.

Further parameters of the experiment are given in Table~\ref{tab:RK-3}. The values of the physical properties of mercury, such as the kinematic viscosity $\nu$ and the electrical  conductivity $\sigma$ at room temperature (about 18 $^o$C) are taken from ~\cite{kirillov2007thermophysical}. The other physical parameters used in the table are the mean flow velocities $U_o$ in the duct and $U_p$ in the inlet pipe, and the thickness $d_w$ and  electrical conductivity $\sigma_w$ of the walls. 

\begin{table}
\begin{center}
\def~{\hphantom{0}}

\begin{tabular}{lcc}

Parameter & Quantity & Value \\  \hline

Duct inner section & $2a$ x $2a$, mm x mm & $56\times56$ \\ 
Pipe radius & $r$, mm & $3$ \\ 
Characteristic length  & $a$, mm & 28 \\ 
Magnetic induction & $B$, T & $0-1.7$\\ 
Length of uniform MF & $L_B$, m & $\approx 0.9$ \\ 
Duct Reynolds number & $\Ru\equiv U_o a /{\nu}$ & $600 - 7000$ \\ 
Duct Hartmann number & $\Ha \equiv B a\sqrt{\sigma_l/{\rho\nu}}$ & $0 - 1300$\\ 
Duct Stuart number & $\N \equiv \Ha^2/\Ru$ & $0 - 2100$ \\
Pipe Reynolds number & $\Ru_p \equiv U_p r /{\nu} = 24\Ru$ & $14000 - 82000$ \\ 
Pipe Hartmann number & $\Ha_p \equiv B r\sqrt{\sigma_l/{\rho\nu}} = 0.1\Ha$ & $0 - 140$\\ 
Pipe Stuart number & $\N_p \equiv \Ha_p^2/\Ru_p$ & $0 - 1$ \\
Wall conductance ratio & $c \equiv \sigma_{w}d_{w}/\sigma a$ & $\lesssim 0.01$\\ 
\end{tabular}
\caption{\label{tab:RK-3} Physical and non-dimensional parameters of the experiment.}
\end{center}
\end{table}

A comment is in order concerning the value of the wall conductance ratio reported in Table \ref{tab:RK-3}. While correct from the formal point of view, the formula used to define $c$ may provide results quite different from the effective value observed in the experiment.  During the experimental program, it is impossible to directly assess the electrical conductivity of the wall. No special methods are used to achieve good contact between mercury and the stainless steel wall other than vacuuming water vapor. It is well known that oxides and other deposits worsen the electrical contact with the wall and decrease the relative electrical conductivity by one to two orders of magnitude. Therefore, instead of applying the formal definition of $c$ we estimate the range of the possible effective values of the wall conductance ratio $c$ in the experiments as between 0.001 and 0.01. The influence of the wall conductance ratio on the flow is explored via numerical simulations and further discussed by \citet{krasnov2022}.

\begin{figure}
 \begin{center}
\includegraphics[width=0.99\textwidth]{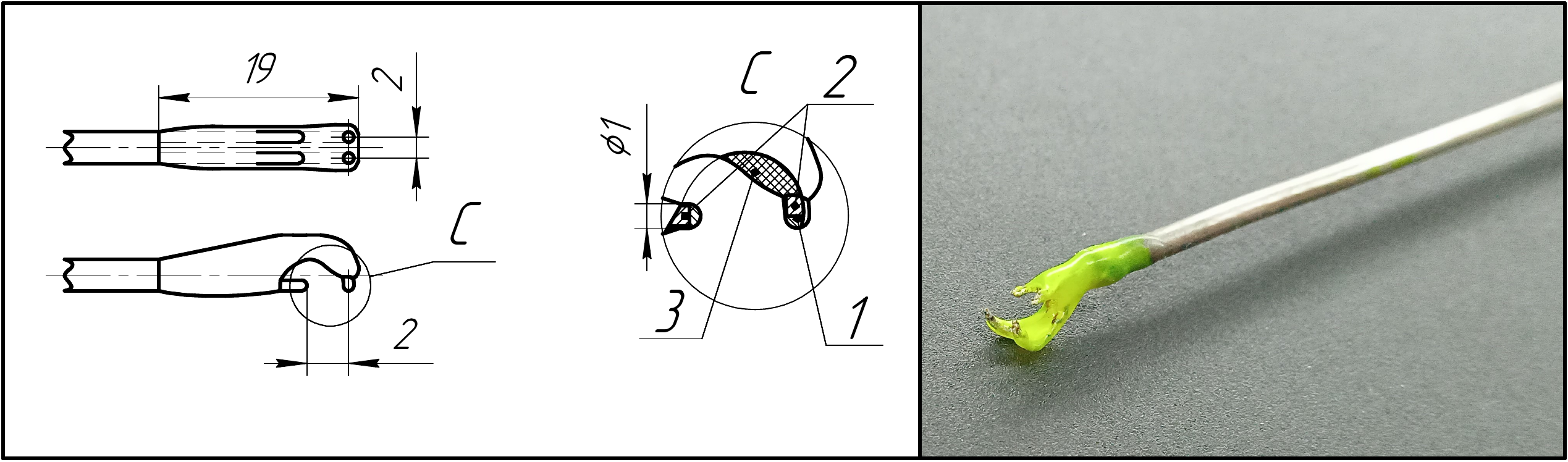}
\begin{tabular}{p{0.1\linewidth}p{0.5\linewidth}p{0.1\linewidth}}
  &  \it{(a)} & \\
\end{tabular}
\includegraphics[width=0.99\textwidth]{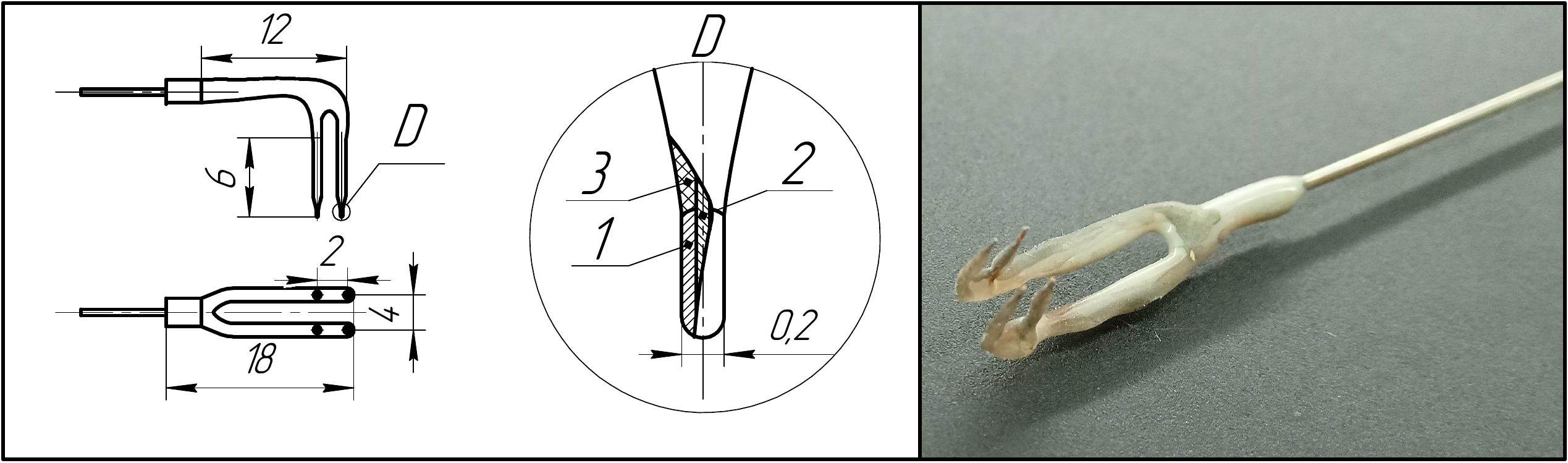}
\begin{tabular}{p{0.1\linewidth}p{0.5\linewidth}p{0.1\linewidth}}
  &  \it{(b)} & \\
\end{tabular}
\includegraphics[width=0.78\textwidth]{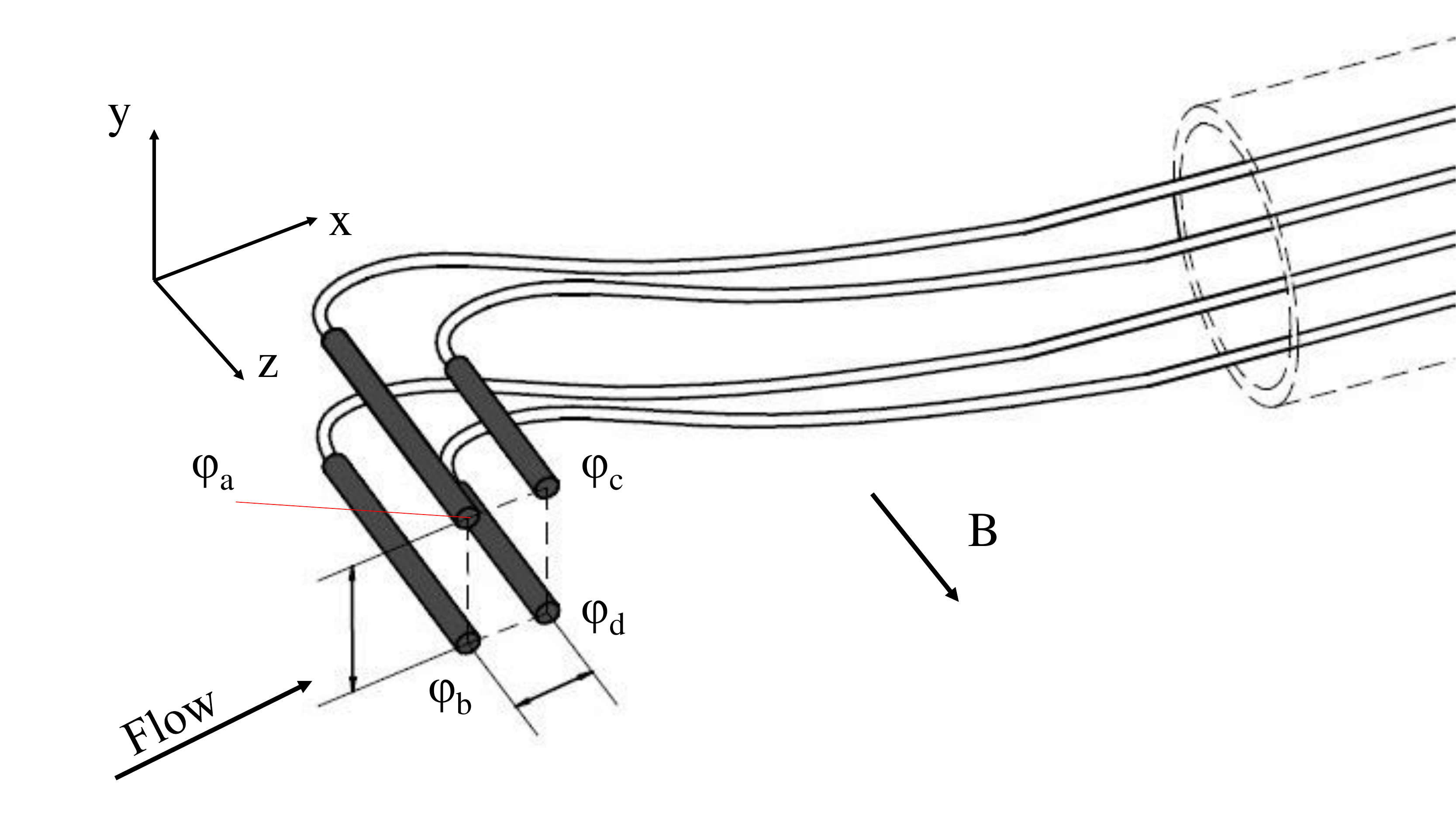}
\begin{minipage}[h]{1\linewidth}
\begin{tabular}{p{0.1\linewidth}p{0.5\linewidth}p{0.1\linewidth}}
\raggedleft \it{(c)} \\
\end{tabular}
\end{minipage}
\end{center}
\caption{ Sensors of different construction: a - Type 1; b - Type 2 c - Schematic representation of the principle of measurements. 1 -- tin contact layer; 2 -- copper electrode; 3 -- insulating layer of epoxy.}
\label{fig:Sensors}
\end{figure}

Flow velocity is measured using the electric potential difference method \citep[see, e. g.,][]{mistrangelo2010perturbing}. A 4-electrode potential sensor illustrated schematically in figure \ref{fig:Sensors}c can measure two components of velocity at a point. With such a sensor, it is possible to simultaneously measure the electric potential $\varphi$ at four electrodes, and then to calculate the local velocity components using the projection of Ohm`s law onto the $x-$ and $y-$ axes. Neglecting electric currents, as their magnitude is small in ducts with poorly conducting walls, and replacing the partial derivatives with their difference approximations gives:
\begin{equation}
u_x\approx\frac{\varphi_{a}-\varphi_{b}}{c_{1} B_0 \Delta l_{y}}\approx\frac{\varphi_{c}-\varphi_{d}}{c_{1} B_0 \Delta l_{y}},
\end{equation}

\begin{equation}
u_y\approx\frac{\varphi_{c}-\varphi_{a}}{c_{1} B_0 \Delta l_{x}}\approx\frac{\varphi_{b}-\varphi_{d}}{c_{1} B_0 \Delta l_{x}},
\end{equation}
where  $\varphi$  is the electric potential at the electrodes, $\Delta l_{y}= 2$mm,  $\Delta l_{x}= 2-4$mm are the distances between the corresponding electrodes, and $c_1$ is the coefficient depending on the conditions of closing the induced currents. For the Hartmann number values used in the experiment, $c_1\approx 1$.

Sensors of two principal designs were  used in the majority of the reported experiments. The sensor of type 1 (see figure \ref{fig:Sensors}a) was used during the first stage of the experimental program. Its design was primarily motivated by the desire to avoid damage of the sensor by a strong jet flow. The sensor of type 2 (see figure \ref{fig:Sensors}b) was built and utilized at later stages of the program in order to validate the results obtained with the sensor of type 1 and to minimize the potential influence of the sensor on measurement data obtained in situations with reversed flow. Yet another, modified sensor of design similar to type 2 was utilized to investigate the jet development far downstream (in the cross-section located at the distance $5.2a$ from jet inlet). 

\begin{figure}
\begin{center}
\includegraphics[width=1.0\textwidth]{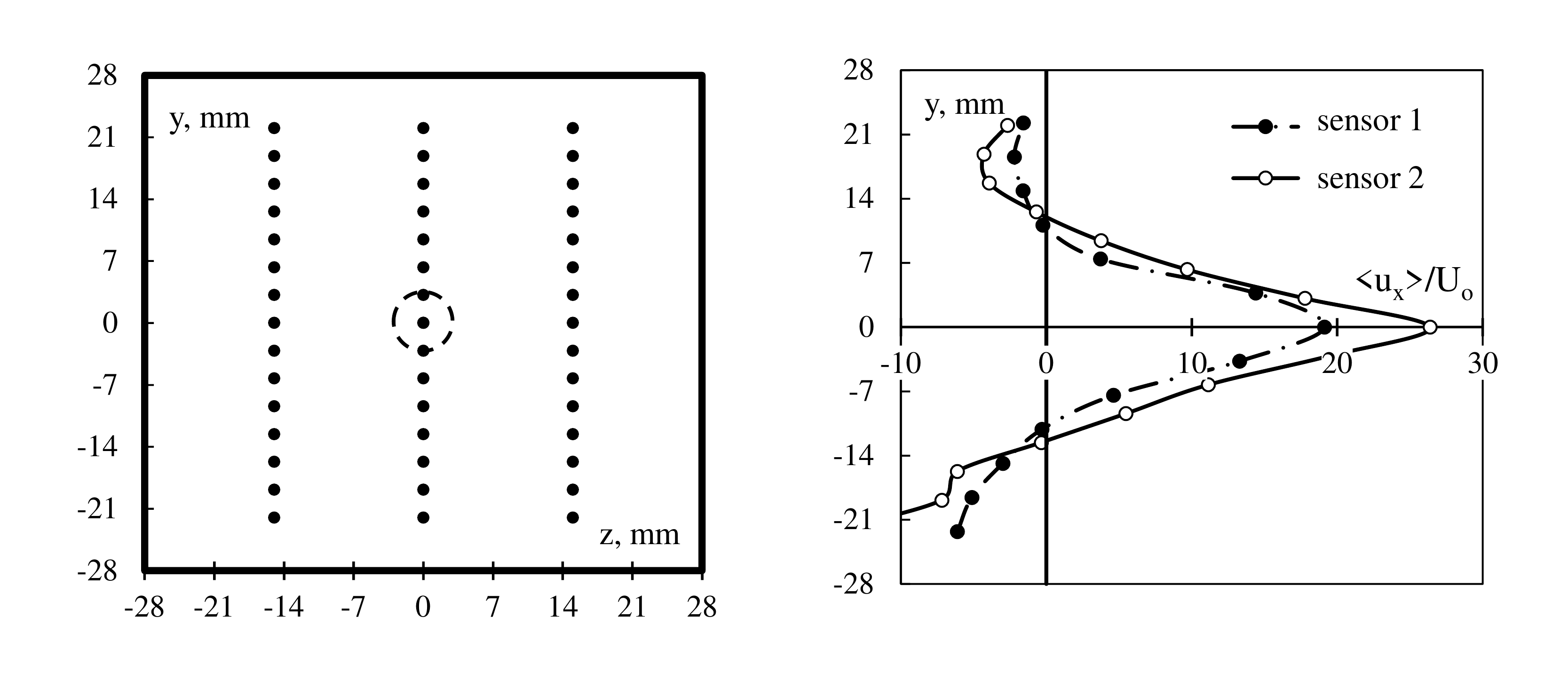}
\begin{minipage}[h]{1\linewidth}
\begin{tabular}{p{0.3\linewidth}p{0.3\linewidth}p{0.30\linewidth}}
\raggedleft \it{(a)} &  & \raggedright \it{(b)}\\
\end{tabular}
\end{minipage}
\end{center}
\caption{(a) Measurement points used to map distributions of time-averaged velocity in duct's cross-section. The inlet shape of the jet is shown by a circle. (b) Profiles of time-averaged  streamwise velocity measured by sensors of type 1 and type 2 at $x=44\textrm{mm}=1.6a$, $z=0$ for \Ru=800$, \Ru_p=19\cdot10^3$, $Ha=200$. Time averaging is over about 600 s.
}
\label{fig:SectionScheme}
\end{figure}

Consistently good agreement between measurements obtained using different sensors was found in all the experiments. An illustration of that is provided in figure \ref{fig:SectionScheme}b. The observed small deviations are attributed to uncertainty of sensor's positioning.  Most of the results reported in this paper are based on the data obtained with the sensor of type 2, except for the diagram in the last figure, where the results of  the experiments performed using sensors of all three types are summarised.

The electrode size of 0.3 mm was chosen to avoid possible errors caused by sensor damage during measurements near walls. Sensors were installed on scanning probes extended in the direction opposite to the flow. The probes allowed us to either record the velocity signal at one location over time or map the mean velocity in a cross-section of the duct by taking time-averaged measurements point by point (see figure \ref{fig:SectionScheme}a for the measurement points utilized for the mapping). Simultaneous measurement of velocities at more than one point was impossible.

\begin{figure}
\centering
\includegraphics[width=1.0\textwidth]{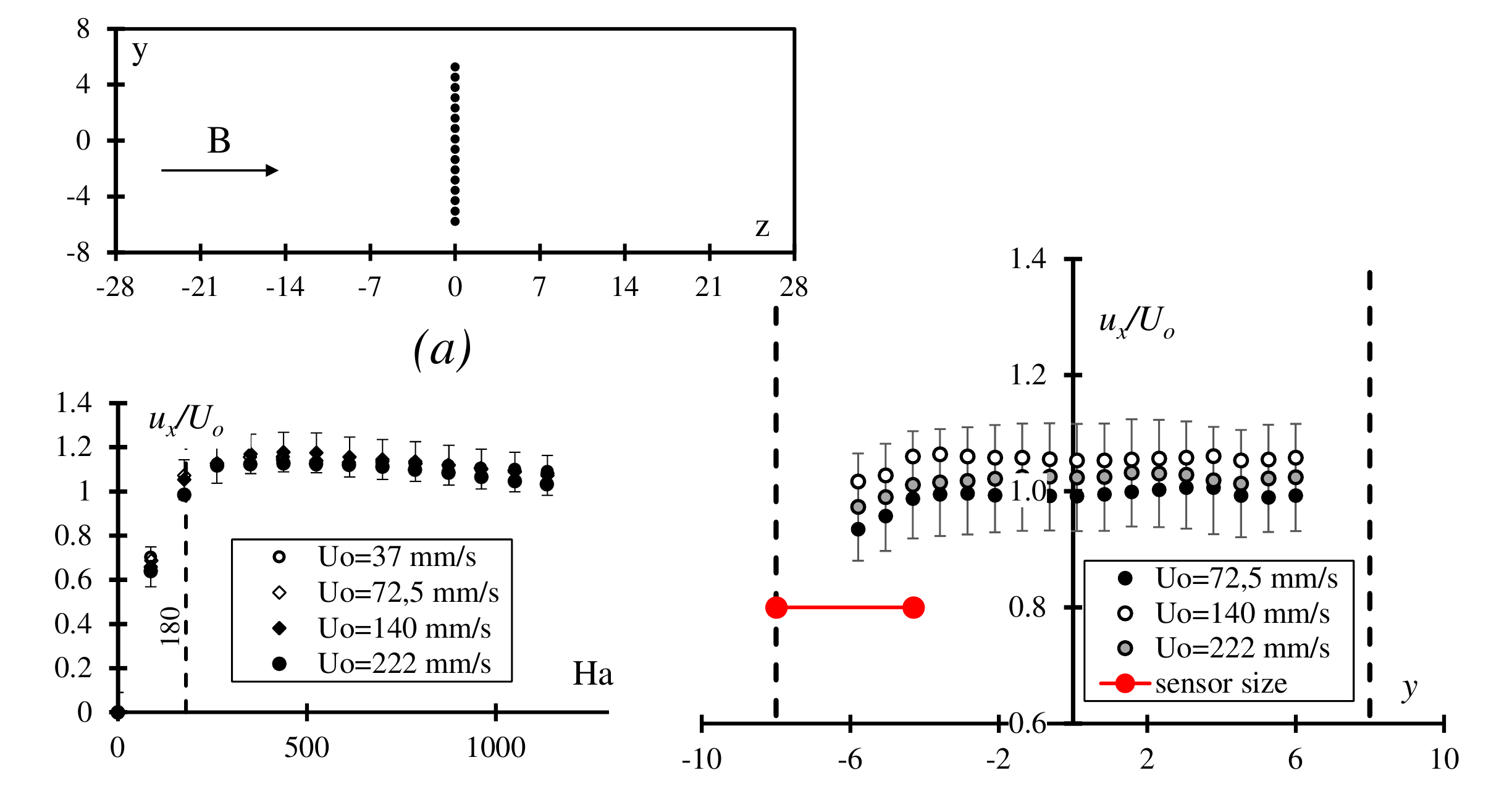}
\begin{minipage}[h]{1\linewidth}
\begin{tabular}{p{0.3\linewidth}p{0.3\linewidth}p{0.30\linewidth}}
\raggedleft \it{(b)} &  & \raggedright \it{(c)}\\
\end{tabular}
\end{minipage}
\caption{Sensor calibration using flow in a uniform duct of rectangular cross-section 56x16 mm. The range of mean velocities $U_0$ and the magnetic field strengths correspond to the values of the Reynolds and Hartmann numbers based on $U_0$ and the half of the longer side of the duct between 9000 and 55000 and between 80 and 1150, respectively. (a) - Duct configuration and points at which velocity profile is measured; (b) - Velocity values at the duct center versus magnetic field; (c) - Velocity profiles at different flow rates and  Ha = 1100. Visible distortion at $-6< y -4$ mm is caused by deformation of the probe from wall contact.} 
\label{fig:Calibration}
\end{figure}

Time-averaged streamwise velocity  values, $<u_x>$ and $<u_y>$, and standard deviations  $\sigma_x$  and $\sigma_y$ are used in further analysis. Standard deviations are calculated as:
\begin{equation}
\sigma_x=\sqrt{\sum_{j=0}^{n-1} \frac{ \left({u_x}_{j}-U_x\right)^{2}}{n-1}}, \quad \sigma_y=\sqrt{\sum_{j=0}^{n-1} \frac{ \left({u_y}_{j}-U_y\right)^{2}}{n-1}},
\label{STD}
\end{equation}
where $n$ is the number of measured points in the velocity signal.

The inherent limitation of the electric potential sensors is their inability to measure velocity in boundary layers. This is due to the facts that the thickness of MHD boundary layers is significantly smaller than the distance between the electrodes, and that positioning the sensor very close to the wall is impossible because of the danger of damaging the sensor and increased uncertainty of measurements due to possible wall-sensor contact.

In the rest of this section we discuss the factors potentially negatively affecting the accuracy of electric potential sensors. We also list measures taken during the experiments to identify and minimize the effect of these factors. It must be stressed that a consistent moderate shift in measured absolute values of velocity is not detrimental for achieving the goals of our study. As we will see below, the flow transformation caused by the magnetic field is described in terms of relative changes of velocity profiles or in terms of centered statistical moments of velocity fluctuations. These characteristics are insensitive to a shift of absolute velocity values, provided the shift is the same in all the experiments.

One potentially negative factor is the thermoelectrical effects associated with the contact between dissimilar metals in a non-uniform temperature field. To minimize it, the mercury loop was water-cooled, so that temperature remained constant and uniform at 291 degrees K. Variations of temperature in the test section were within 0.4 K.

Another factor is the error (zero shift) caused by electromagnetic noise in measurement equipment. To avoid this effect, each experimental session started with measurements performed at zero imposed magnetic field. The registered small deviation from the expected zero value of velocity was then used for correction of measurements in production runs performed at non-zero imposed magnetic field. Furthermore, two acquisition systems were used, one based on an NI cDAQ 9213 measuring module with 100Hz antialiasing analog filtering, and another on NI 4071 digital multimeters with multi-speed digitization and digital processing of 1kHz signals. After the aforementioned correction, the two systems showed consistent results. 

The biggest potential factor negatively affecting measurement accuracy is the disturbance of the flow by the probe. The disturbance may happen directly, with the probe modifying the velocity as a solid obstacle, or indirectly via MHD effects caused by the flow around the probe. One already mentioned step undertaken in the experiments to reduce the probe's influence was to always insert it in the direction opposite to the mean flow. To evaluate the remaining effects, a special calibration experiment illustrated in figure \ref{fig:Calibration} was carried out. The geometry of the test section was replaced by that of a long straight duct of rectangular cross-section $56\times 16$ mm. A uniform magnetic field was imposed along the $z$-axis. Time-averaged streamwise velocity $\langle u_x\rangle$ was measured along the $y$-axis in the middle of the duct at the distance of 600 mm downstream from the entrance into the magnetic field. It is anticipated in such a system that a moderate magnetic field would produce a flow with experimentally detectable sidewall and the centerline  value of $\langle u_x\rangle$ exceeding the mean velocity $U_0$. At stronger magnetic fields, the sidewall boundary layers would be undetectable thin, and the measured  velocity would show a nearly flat profile along the $y$-axis, with $\langle u_x\rangle \sim U_0$. The results of experiments shown in figure \ref{fig:Calibration} agree with the expectations. An exception is flows with weak magnetic field (approximately $\Ha$ below 180), in which case accuracy of the sensor is expected to be low. Figure \ref{fig:Calibration} illustrates the agreement by the variation of the ratio $\langle u_x\rangle/U_0$ with the magnetic field strength (in figure \ref{fig:Calibration}b) and by full measured velocity profiles at a strong magnetic field (in figure \ref{fig:Calibration}c). We conclude that the accuracy of the sensor at all but the smallest Hartmann numbers considered in our experiments is confirmed by the test. 

The size of the sensor does not allow us to measure velocity distribution within the boundary layers and, therefore, unambiguously identify the velocity profile near the walls. It is, however, visible, that the velocity in figure \ref{fig:Calibration}c does not show the  tendency toward an M-shaped profile typical for a flow in a duct with thin electrically conducting walls \citep{muller2001magnetofluiddynamics}. This indicates a very low value of effective conductance rate $c$ consistent with our estimates in Table \ref{tab:RK-3}. 

All the results reported in the following sections of the paper satisfy three criteria: \emph{(i)} measurements of the time-averaged values of $u_x$ are consistent with $U_0$ measured by an electromagnetic flowmeter,  \emph{(ii)} nearly identical results are obtained in experiments performed with different acquisition systems, and  \emph{(iii)} measurements are performed at magnetic fields high enough to ensure the stable performance of the sensor demonstrated in calibration experiments.  
The experiments, in which these conditions are not satisfied (this happens, for example in cases with low magnetic field) are considered insufficiently accurate and not reported.

\section{Results}
\label{sec:sample:experimentexpl}
Significant unsteadiness of the flow, with high-amplitude fluctuations of velocity recorded downstream of the jet inlet, is found in all the experiments reported in this paper. It must be stressed that the unsteadiness cannot be a result of transition to turbulence within the duct itself. As first identified by \cite{murgatroyd1953cxlii} and later  discussed on the basis of experimental data by \citet{branover1978magnetohydrodynamic} and from the theoretical perspective by \citet{Zikanov:2014}, a transition or laminarization occurs in laboratory flows when $\Ru/\Ha$ traverses the range between 200 and 400.  All the experiments discussed below have smaller values of $\Ru/\Ha$. The ratio $\Ru_p/Ha_p=240 \Ru/\Ha$ determining the possibility of transition within the inlet pipe can be in or above the transition range, so some turbulent fluctuations of moderate amplitude are possible at the jet inlet. We conclude that the high-amplitude velocity fluctuations recorded in the experiments can only be results of the instability of the jet itself, and that the instability is likely to be affected by perturbations at the jet's inlet.

Presence of high-amplitude fluctuations creates a challenge for experimental characterization of the flow. In particular, time series of long duration (up to 600 seconds) are required to accurately measure time-averaged velocity at each point.

Measurements are made in the cross-section $x=44$ mm $=1.6a$, chosen so that the jet has already developed its unsteady structure resulting from the instability but has not strongly decayed yet. A distribution of time-averaged velocity in the duct's cross-section is found by positioning  the sensors along the lines $z=0$, $z=-a/2$, and $z=a/2$ perpendicular to the magnetic field, i.e. from one sidewall to another (see figure \ref{fig:SectionScheme}a) and recording a full time series  at each measurement point. 

\begin{figure}
\centering
\includegraphics[width=1.0\textwidth]{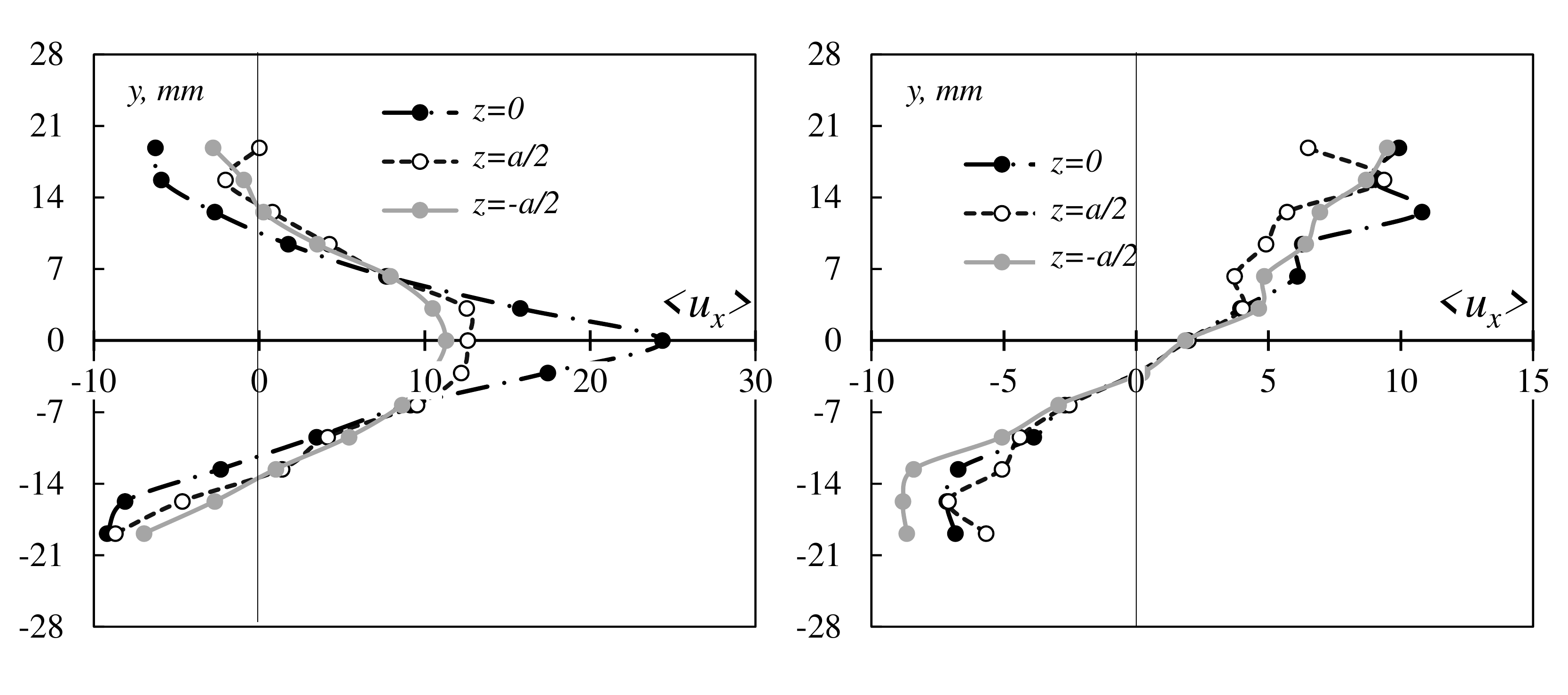}
\begin{minipage}[h]{1\linewidth}
\begin{tabular}{p{0.3\linewidth}p{0.3\linewidth}p{0.30\linewidth}}
\raggedleft \it{(a)} &  & \raggedright \it{(b)}\\
\end{tabular}
\end{minipage}
\caption{An illustration of two-dimensionalization of the velocity field caused by magnetic field. Time-averaged longitudinal velocity profiles at $x=44$ mm $=1.6a$ are shown for three values of $z$ in the flow at $\Ru=800$, $\Ru_p=19\cdot10^3$. $\Ha=200$ in (a). $\Ha=1300$ in (b).}

\label{fig:FirstVelocityProfile}
\end{figure}

\begin{figure}
\centering
\includegraphics[width=1.0\textwidth]{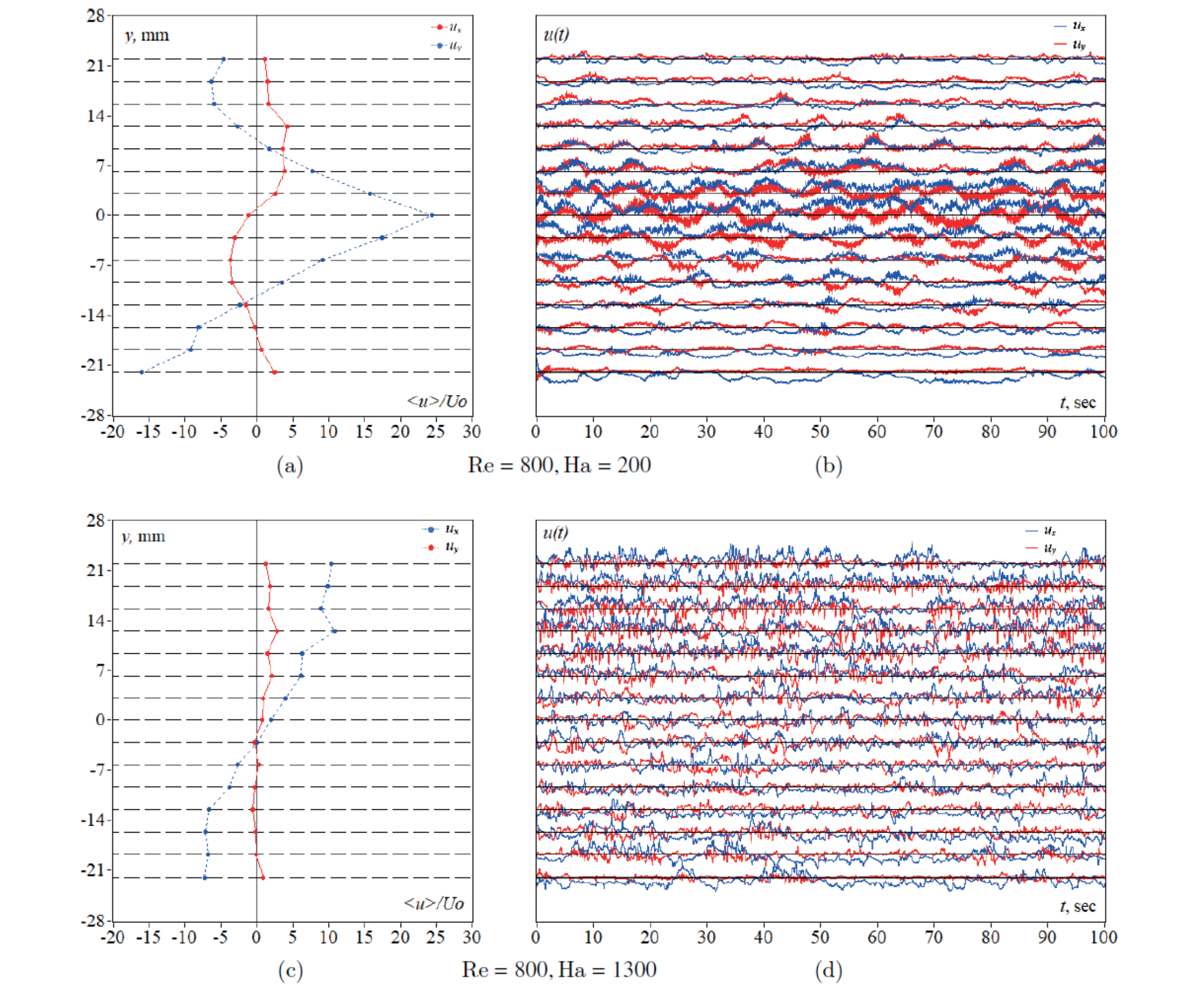}


\caption{
Longitudinal $u_x$ (blue curves) and transverse $u_y$ (red curves) velocity components measured  at points located along the line $z=0$ at $x=1.6 a=44mm$ in flows with $\Ru=800, \Ha=200$ ((a) and (b)) and  $\Ru=800,  \Ha=1300$ ((c) and (d)). Profiles of $\langle u_x \rangle$ and $\langle u_y \rangle$ obtained by averaging over 610 s are shown in (a) and (c). Velocity fluctuations around the time-averaged values are shown in (b) and (d). In the latter, each pair of red and blue curves shows a velocity signal measured at a $y$-location indicated by a respective point in the left-hand side graph. The velocity scale in (b) and (d) is such that the half-distance between the black lines corresponds to $60U_0$.
}
\label{fig:UxUy}
\end{figure}

\begin{figure}
\centering
\includegraphics[width=0.9\textwidth]{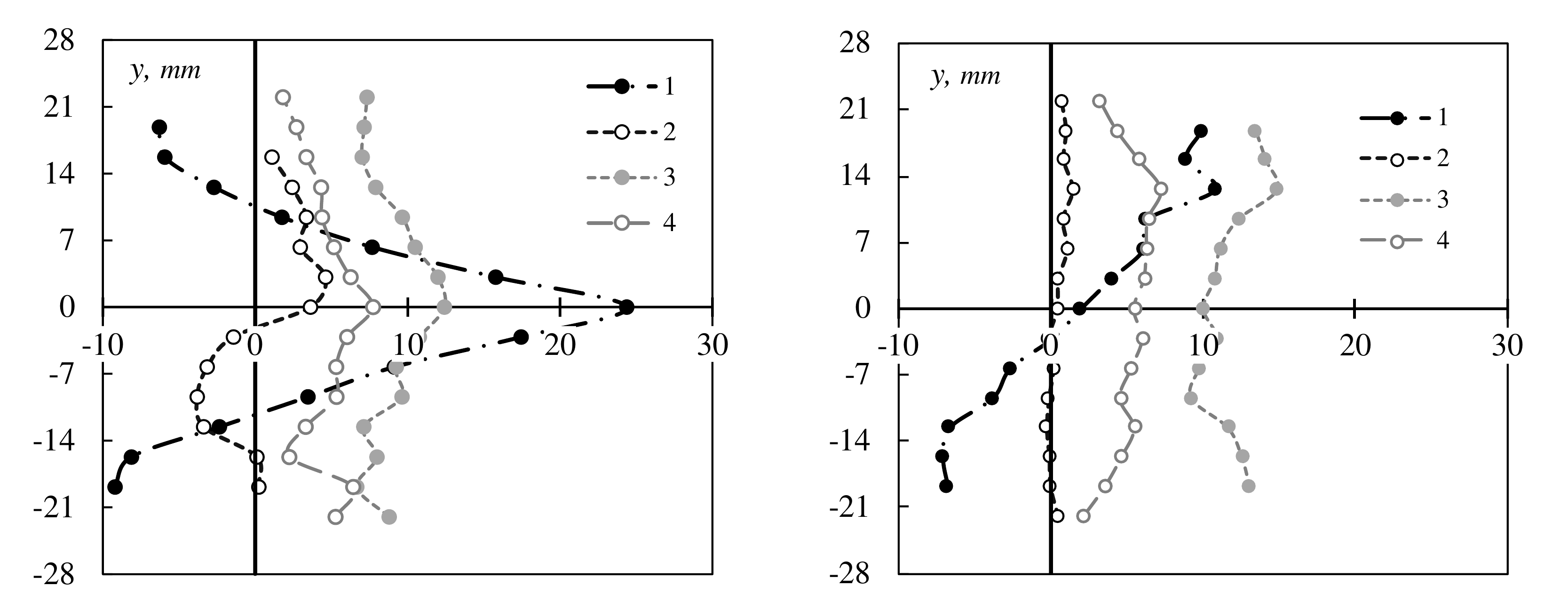}
\begin{minipage}[h]{1\linewidth}
\begin{tabular}{p{0.3\linewidth}p{0.3\linewidth}p{0.30\linewidth}}
\raggedleft \it{(a)} &  & \raggedright \it{(b)}\\
\end{tabular}
\end{minipage}
\includegraphics[width=0.9\textwidth]{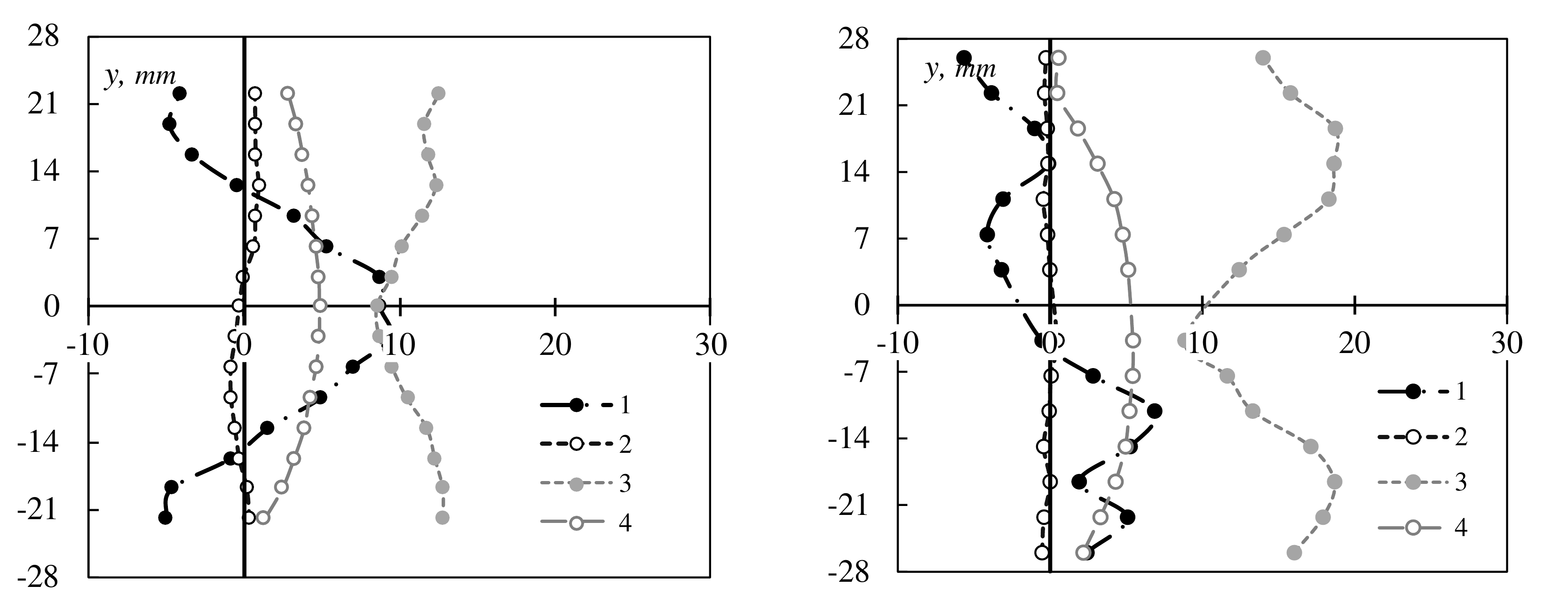} 
\begin{minipage}[h]{1\linewidth}
\begin{tabular}{p{0.3\linewidth}p{0.3\linewidth}p{0.30\linewidth}}
\raggedleft \it{(c)} &  & \raggedright \it{(d)}\\
\end{tabular}
\end{minipage}
\caption{Time-averaged characteristics of flows at (a) $\Ha=200$, $\Ru=800$;  (b) $\Ha=1300$, $\Ru=800$; (c) $\Ha=550$, $\Ru=5000$; and (d) $\Ha=1300$, $\Ru=3000$. Profiles measured at $x=1.6a$, $z=0$ are shown. 
1 - $\langle u_x \rangle/U_0$; 2 - $\langle u_y \rangle/U_0$; 3 - $\sigma_x/U_0$; 4 - $\sigma_y/U_0$.
} 
\label{fig:VelocityProfile}
\end{figure}

The results for the time-averaged velocity components $\langle u_{x}\rangle$ and $\langle u_{y}\rangle$ are presented in figures \ref{fig:SectionScheme}, \ref{fig:FirstVelocityProfile}, \ref{fig:UxUy}  and \ref{fig:VelocityProfile}. The tendency towards formation of quasi-two-dimensionality in an initially round jet is illustrated in figure \ref{fig:FirstVelocityProfile} by velocity profiles at $\Ru=800$. At $Ha=200$ there is already strong flow at $z=\pm a/2$, but significant difference between the profile at the center of the duct $z=0$ and the profiles measured at $z=-a/2$ and $a/2$ is visible (see figure \ref{fig:FirstVelocityProfile}a).  At $Ha=1300$, the three profiles are close to each other demonstrating quasi-two-dimensionality (see figure \ref{fig:FirstVelocityProfile}b). 

Figure \ref{fig:FirstVelocityProfile} also illustrates existence of two clearly different states of the flow. The time-averaged velocity profiles are symmetric with respect to the centerline $y=0$ at low Hartmann numbers, such as $\Ha=200$ in figure \ref{fig:FirstVelocityProfile}a. Strong asymmetry is observed at higher $\Ha$, such as $Ha=1300$ in figure \ref{fig:FirstVelocityProfile}b.

The two typical behaviours observed at moderate and high values of $\Ha$ are further illustrated in figure \ref{fig:UxUy} on the example of flows at $\Ru=800$, $\Ha=200$ and $\Ru=800$, $\Ha=1300$. Time-averaged values and waveforms of longitudinal $u_x$ and transverse $u_y$ velocity components  measured at points distributed along $y$ at $z=0$ are provided. The waveforms show instantaneous simultaneously measured local values of $u_x$ and $u_y$, from which time-averaged values computed at the same $y$ are subtracted. One should keep in mind that waveforms corresponding to different measurement points are recorded during different periods of time, so any apparent correlation between them is purely coincidental.

Figures \ref{fig:UxUy}a,b illustrate a typical flow observed at moderate values of $\Ha$. The profile of $\langle u_x \rangle$ is symmetric. It has a  peak at $y=0$ and shows positive values of $\langle u_x \rangle$ at about $-10<y<10$ mm. This area coincides with the area of high-amplitude fluctuations in the waveforms. We conclude that the interval $-10<y<10$ mm corresponds to the zone swept by an oscillating central jet. Outside this interval, i.e. at $y<-10$ mm or $y>10$ mm, $\langle u_x \rangle<0$ and the amplitude of velocity fluctuations is significantly lower. These areas are identified as those of a weakly turbulent reversed flow. The picture of an expanding central jet is further confirmed by the profile of  $\langle u_y \rangle$ shown in figure \ref{fig:UxUy}a. After a correction for a zero shift we observe a flow in the positive $y$-direction at $y>0$ and in the negative $y$-direction at $y>0$. 

The waveforms in figure \ref{fig:UxUy}b show strong fluctuations in the central area, especially at $y=0$. The maximum amplitude exceeds $60U_0$ at some moments. The dominant frequencies correspond to the time period of about 6-7 s. There are also oscillations of much higher frequency and lower amplitude corresponding to three-dimensional turbulence. Another interesting feature is the correlation between fluctuations of $u_x$ and $u_y$ observed in the side regions of the jet, but not in its center. Positive peaks of $u_x$ correlate with positive or negative peaks of $u_y$ at, respectively, $y>0$ or $y<0$. Such events are identifiable as excursions of the jet into the areas of positive or negative $y$. 

\begin{figure}
\centering
\includegraphics[width=0.99\textwidth]{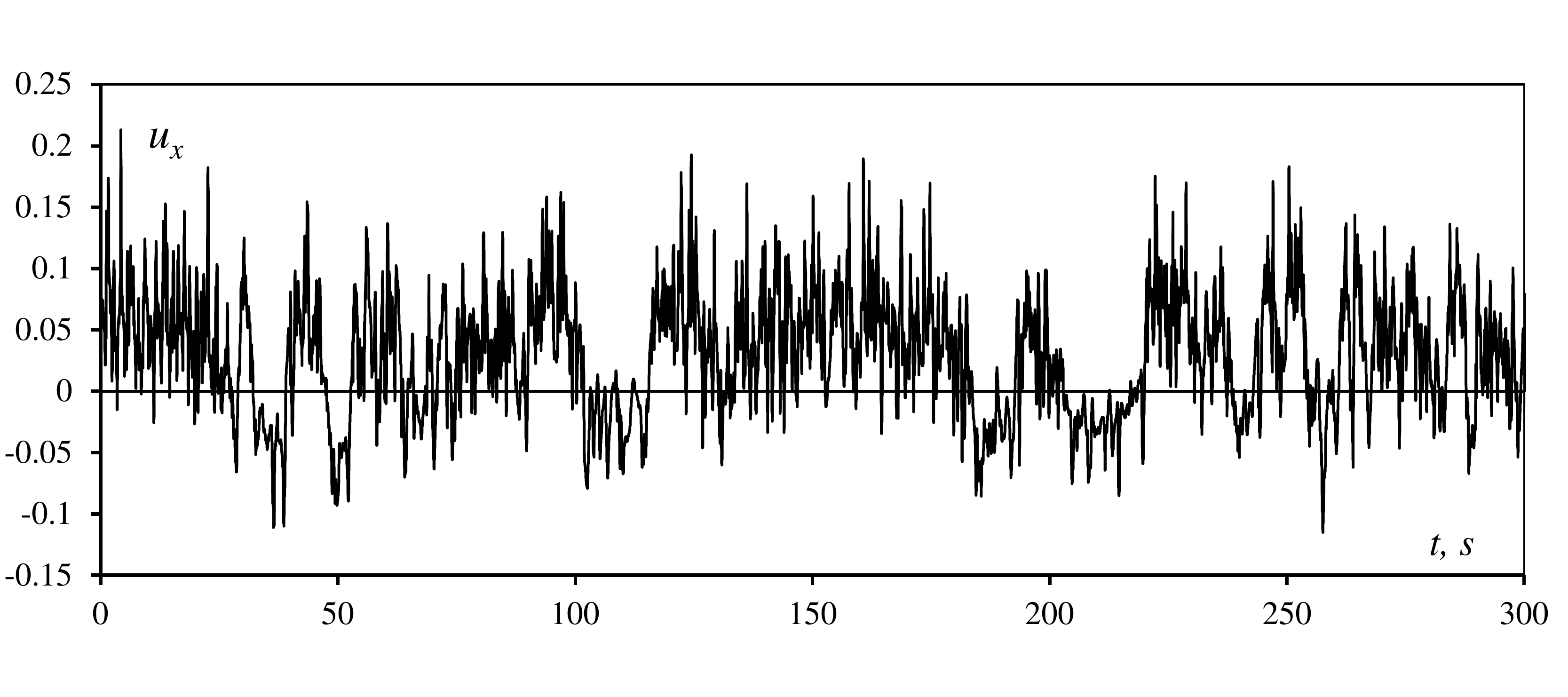}

\caption{Signal of longitudinal velocity $u_x$ measured at $z=0$, $y=a/2$ in the flow with $\Ru=900$, $\Ha=1300$} 
\label{fig:Intermitency}
\end{figure}

An entirely different structure of the flow is presented by the measurements made at $\Ru=800$, $\Ha=1300$ (see figures \ref{fig:UxUy}c,d). The asymmetry, which we have already presented in figure \ref{fig:FirstVelocityProfile}, demonstrates that the jet is shifted toward one side of the duct. It must be stressed that the apparent shift is a result of time-averaging. The  signal of $u_x$ at $y=a/2$ shown in figure \ref{fig:Intermitency} illustrates the fact that velocity fluctuates and may become negative at some moments. Once time-averaging over a sufficiently long time period is taken, however, the jet is found consistently located on one side of the duct, at positive (as in figures \ref{fig:VelocityProfile}b and \ref{fig:UxUy}c) or negative $y$. No preference for one side and no spontaneous switches of the time-averaged profiles from one side to another have been found in the experiments. A possibility of such behaviours cannot be excluded, since no special studies that could detect them (statistical studies with multiple realizations of the same flow or very long measurement runs) have been completed.

The profiles of $\langle u_x \rangle$ and $\langle u_y \rangle$ in figure \ref{fig:UxUy}c also show a reversed flow in the negative $y$ part of the duct. It is nearly as strong as the jet itself. The global structure of the asymmetric flow structure can be characterized as that of a macrovortex. 

The velocity waveforms in figure \ref{fig:UxUy}d demonstrate significant fluctuations in the entire measurement region. The typical amplitude and frequency of the fluctuations are higher in the zone of the jet at positive $y$ than in the zone of reversed flow. It can also be seen that high-frequency turbulence oscillations present in the flow at  $\Ha=200$ (see figure \ref{fig:UxUy}b) are not seen at $\Ha=1300$. This is attributed to quasi-two-dimensionality of flow structures at such high $\Ha$.

The profiles of mean velocities $\langle u_x \rangle$, $\langle u_y \rangle$ and standard deviations $\sigma_x$, $\sigma_y$ (see (\ref{STD})) nondimensionalized by the mean duct velocity $U_0$ are shown in figure \ref{fig:VelocityProfile}. They further illustrate existence of the two types of flow structure found in the experiments: a central jet and a macrovortex. The structure of the macrovortex becomes more complex at high $\Ru$ as illustrated by multiple extrema of the profile of $\langle u_x \rangle$ in figure \ref{fig:VelocityProfile}d. 

\begin{figure}
\centering
\includegraphics[width=1\textwidth]{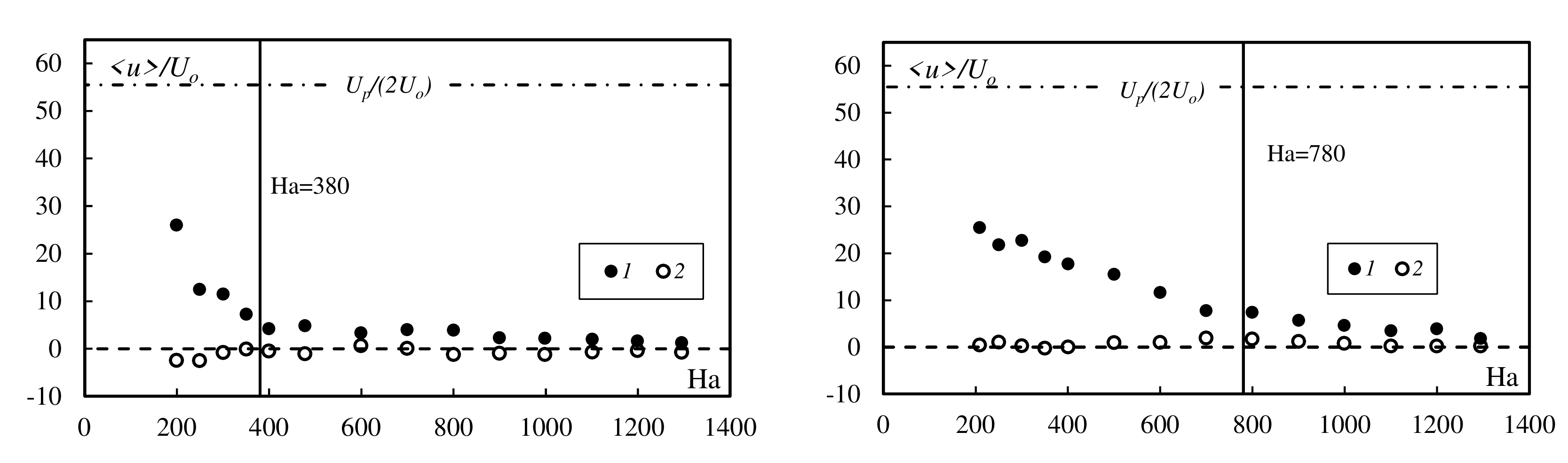}
\begin{minipage}[h]{1\linewidth}
\begin{tabular}{p{0.21\linewidth}p{0.43\linewidth}p{0.30\linewidth}}
\raggedleft \it{(a)} &  & \raggedright \it{(b)}\\
\end{tabular}
\end{minipage}
\includegraphics[width=1\textwidth]{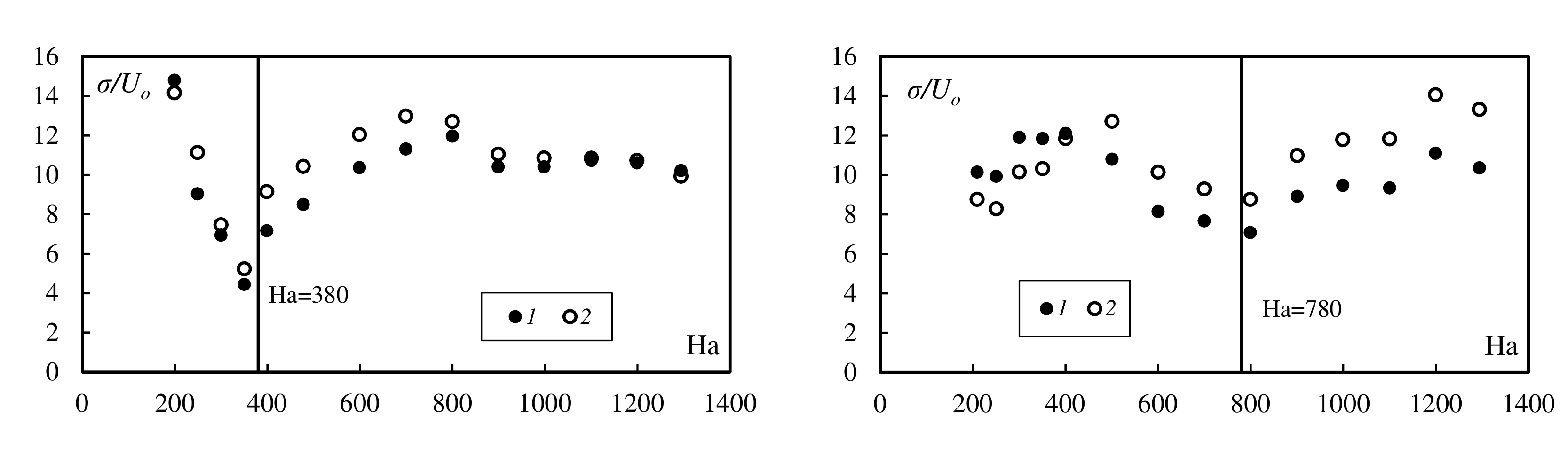}
\begin{minipage}[h]{1\linewidth}
\begin{tabular}{p{0.21\linewidth}p{0.43\linewidth}p{0.30\linewidth}}
\raggedleft \it{(c)} &  & \raggedright \it{(d)}\\
\end{tabular}
\end{minipage}
\caption{Velocity measurements at the center of the duct $y=z=0$ in the cross-section $x=1.6a$. Time-averaged velocity components (a,c) and standard deviations (b,d)  based on signals of length 600 s are shown. $\Ru=800$ (a,c) and $\Ru=3000$ (b,d). 1-- $\langle u_x \rangle$ and $\sigma_x$. 2 -- $\langle u_y \rangle$ and $\sigma_y$. The horizontal dash-dot line in (a) and (b) shows the value equal to a half of the mean inlet velocity of the jet. The vertical solid lines indicate $\Ha_{cr}$ separating central jet and macrovortex regimes (see text).  } 
\label{fig:Ha_increase}
\end{figure}

It is possible to differentiate between flow states with a central jet and a macrovortex using magnitudes of the time-averaged velocity components and their standard deviations at the central point $y=z=0$. Results of such an analysis carried out for two values of $\Ru$ and the entire explored range of $\Ha$ are presented in figure \ref{fig:Ha_increase}. To indicate the magnitude of velocity with respect to the mean inlet velocity of the jet $U_p=110U_0$, a horizontal line corresponding to $U_p/2$ is drawn in figure \ref{fig:Ha_increase}a,b.

The data for $\langle u_x \rangle$, $\sigma_x$, and $\sigma_y$ (although not for $\langle u_y \rangle$, which is nearly zero for all $\Ha$) clearly show existence of two intervals of $\Ha$ with distinct behaviours. The critical Hartmann numbers separating the intervals (shown as vertical lines in figure \ref{fig:Ha_increase})  are approximately $\Ha_{cr}=380$ at $\Ru=600$ and  $\Ha_{cr}=780$ at $\Ru=3000$. At $\Ha<\Ha_{cr}$, $\langle u_x \rangle$ is large and gradually decreases from its maximum $\uxa\approx 25U_0$ at $\Ha=200$ (as explained in section \ref{sec:description}, this is the lowest $\Ha$ at which accurate measurements are possible) to nearly zero. The intensity of velocity fluctuations quantified by $\sigma_x$ and $\sigma_y$ also decreases and reaches its minimum at $\Ha=\Ha_{cr}$. 

The interval $\Ha>\Ha_{cr}$ is characterized by small values of $\uxa/U_0$ and by $\sigma_x$ and $\sigma_y$ growing with $\Ha$ to large values (about $12U_0$ to $14U_0$). Some decrease of the standard deviation at very high $\Ha$ is observed at $\Ru=600$, but not, in the explored range of $\Ha$, at $\Ru=1300$. 

A detailed discussion of the flow transformation causing the radical change of behavior is provided in section \ref{sec:Discussion}. Here we only state the evident conclusion that the flow takes the form of a central jet at $\Ha<\Ha_{cr}$ and of a macrovortex at $\Ha>\Ha_{cr}$.

\begin{figure}
\centering
\includegraphics[width=0.99\textwidth]{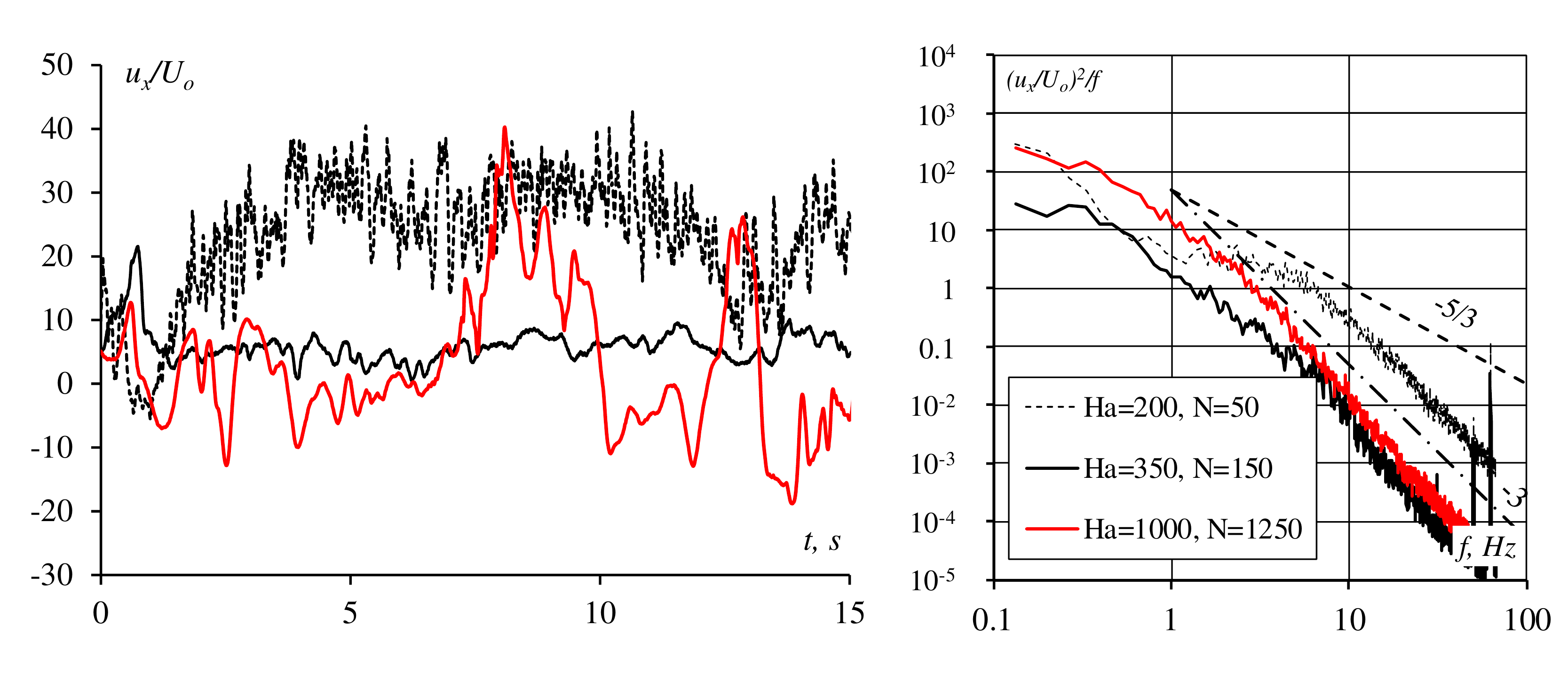}
\begin{minipage}[h]{1\linewidth}
\begin{tabular}{p{0.21\linewidth}p{0.43\linewidth}p{0.30\linewidth}}
\raggedleft \it{(a)} &  & \raggedright \it{(b)}\\
\end{tabular}
\end{minipage}
\includegraphics[width=0.99\textwidth]{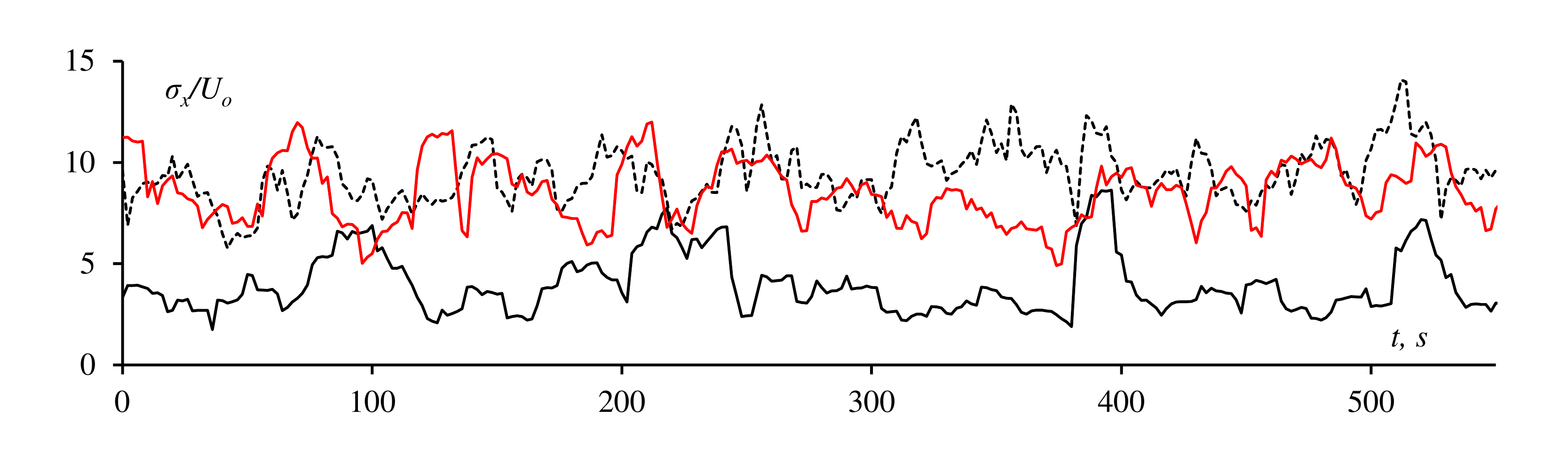}
\begin{minipage}[h]{1\linewidth}
\begin{tabular}{p{0.1\linewidth}p{0.5\linewidth}p{0.1\linewidth}}
  & \centering \it{(c)} & \\
\end{tabular}
\end{minipage}
\caption{Properties of the  signal of longitudinal velocity $u_x$ measured at the center $y=z=0$ of the duct in the cross-section $x=1.6a$ at $\Ru=800$ and $\Ha=200$, 350, and 1000. (a), signals in a 15 s window. (b), energy power spectra based on 600 s of observations. The apparent peaks at $f=50$ Hz are caused by interference from the power supplies. The slopes $\sim f^{-5/3}$ and $\sim f^{-3}$ are shown for guidance. (c), curves of the standard deviation calculated in a moving 15 s window.} 
\label{fig:Re800Waveforms_and_spectra}
\end{figure}

The impact of the magnetic field on the properties of velocity signal measured at the duct's center is illustrated in figure \ref{fig:Re800Waveforms_and_spectra}. Properties of $u_x$ at $\Ru=800$ and $\Ha=200$ (central jet), $\Ha=1000$ (macrovortex), and $\Ha=350$ (close to $\Ha_{cr}$) are shown. We see the typical features of flow transformation observed in other MHD flows in experiments and numerical simulations, e.g., by \citet{kolesnikovTsinob1972,Alemany:1979,Burattini:2010,Verma:2017,Zikanov:2019}. The transition into a quasi-two-dimensional form at growing $\Ha$ is associated with suppression of high-frequency fluctuations, growth of energy of low-frequency fluctuations, and steeper energy spectra with the slope approaching $\sim f^{-3}$ (see figures \ref{fig:Re800Waveforms_and_spectra}a,b).  A feature specific to the jet flow and already visible in figure \ref{fig:Ha_increase}c is illustrated in figure \ref{fig:Re800Waveforms_and_spectra}c. The typical amplitude of velocity oscillations varies non-uniformly with $\Ha$. It decrease from about $10U_0$ to about $5U_0$ when $\Ha$ changes from 200 to 350 and increases back to about $10U_0$ when $\Ha$ increases to 1000.

\begin{figure}
\centering
\includegraphics[width=0.99\textwidth]{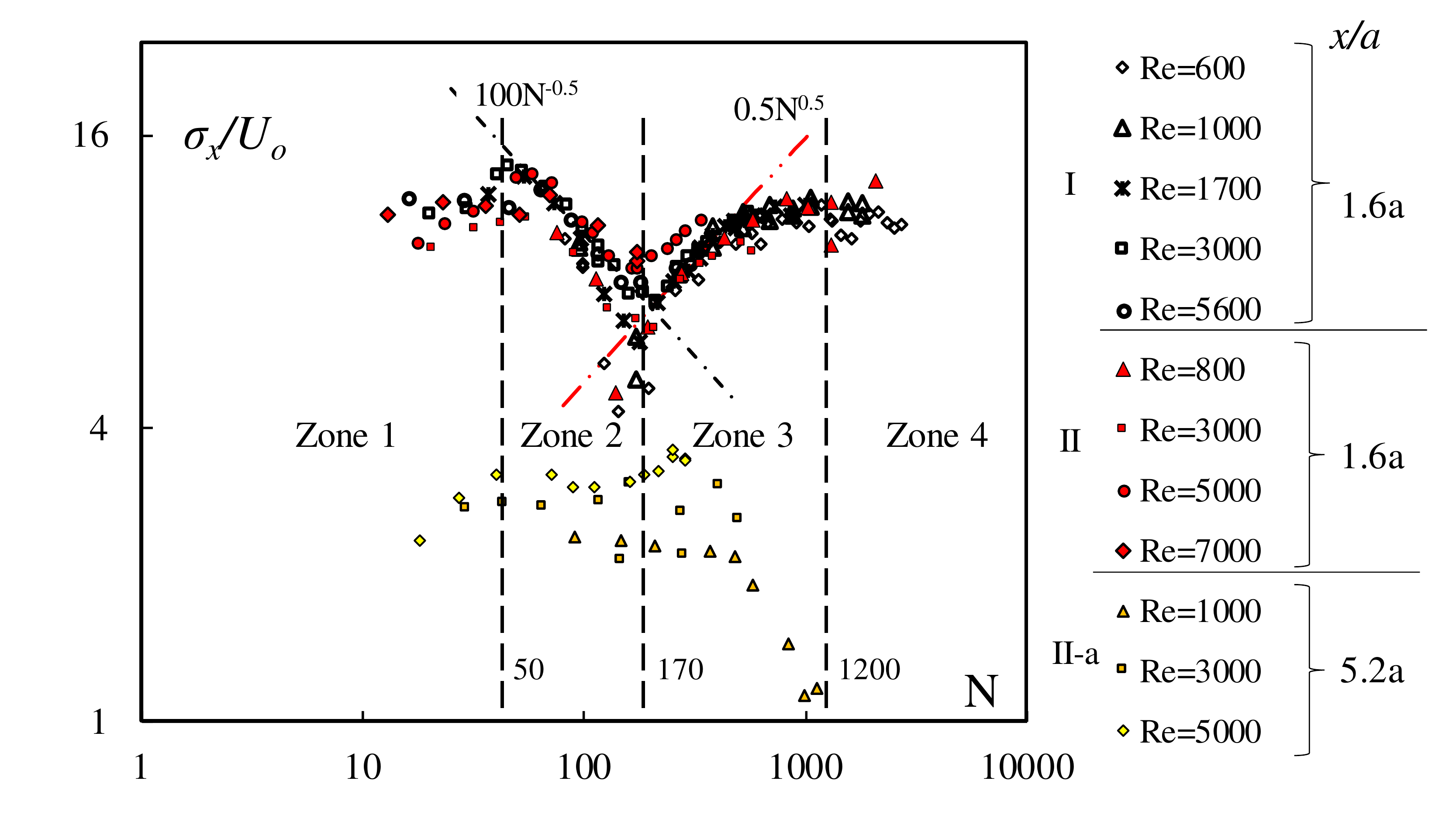}
\textit{(a)}
\includegraphics[width=0.99\textwidth]{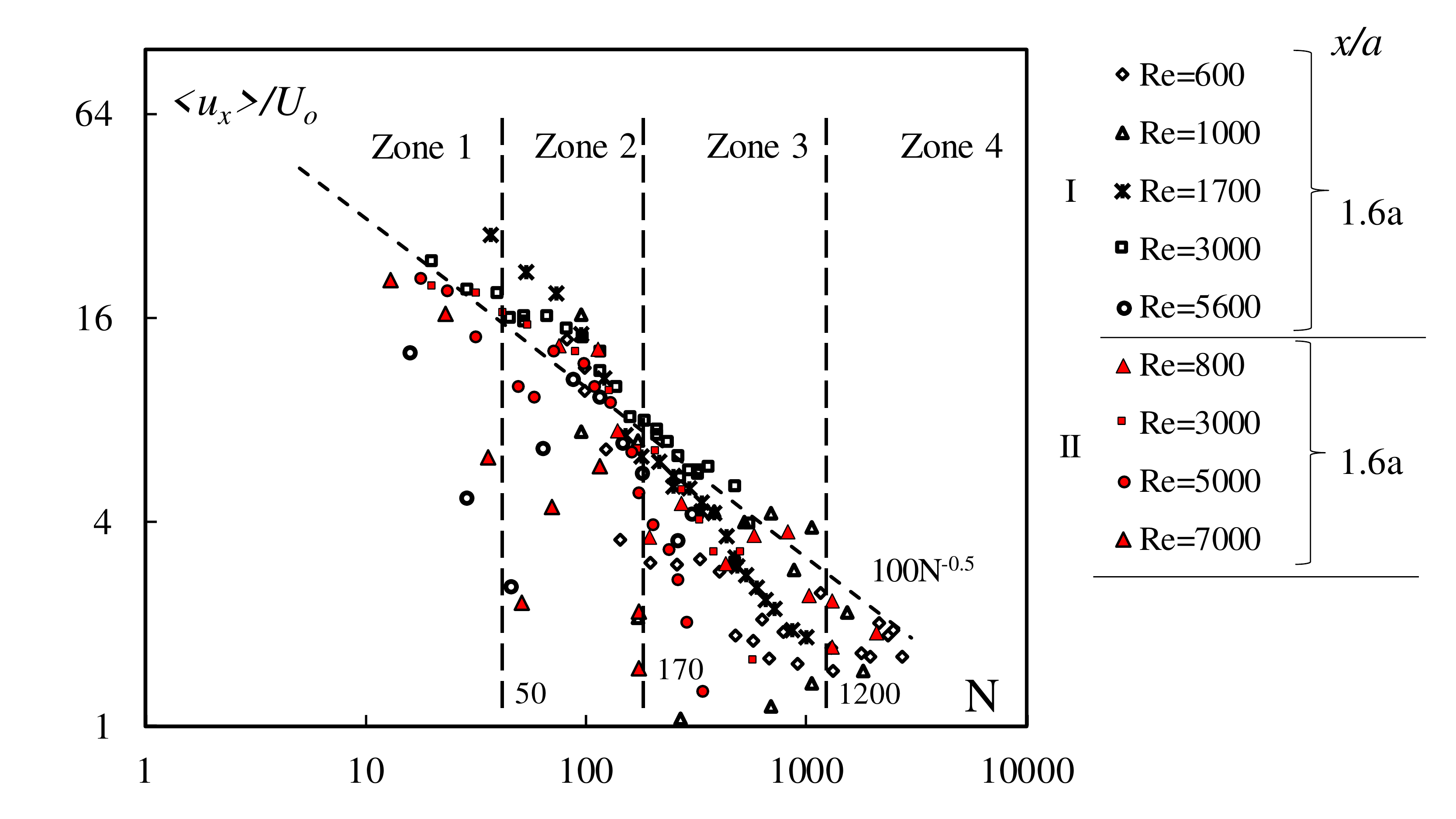}
\textit{(b)}
\caption{ Nondimensionalized standard deviation $\sigma_x$ (a) and mean value $\langle u_x \rangle$ (b) of longitudinal velocity measured at the duct center $y=z=0$ versus the Stuart number $\N$. Data for the cross-sections $x=1.6a$ (I and II) and $x=5.2a$ (II-a) are shown. Measurements series I and II are completed using, respectively, sensors of type 1 and 2 shown in figure \ref{fig:Sensors}a,b. Measurements of series I are completed using sensor of type 3 similar to the sensor of type 2 in its design (see section \ref{sec:description}.}  
\label{fig:N-diagram}
\end{figure}

The results of all the reported measurements of longitudinal velocity at the duct center are summarized in figure \ref{fig:N-diagram}. Standard deviation and mean value of the longitudinal velocity scaled with $U_0$ are shown as functions of $\N$. We see that the two trends presented  by figure  \ref{fig:Ha_increase}c are universal. The mean centerline velocity decreases with the strength of the magnetic field effect and becomes small when the field is strong. The amplitude of the fluctuations has two peaks, one at small and one at large $\N$, and a minimum at intermediate values of $\N$.

The line $\langle u_x\rangle \sim \N^{-0.5}$ in figure \ref{fig:N-diagram}b corresponds to the theoretically determined MHD decay of velocity in a submerged uniform jet in an infinite domain \citep{davidson1995magnetic}. It is not surprising that the experimental data do not reproduce this slope closely. The real jet is spatially evolving rather than uniform and occurs in a bounded domain of the duct.

We see in figure \ref{fig:N-diagram}a that all the data collected in the cross-section $x=1.6a$ at various values of $\Ru$ and using sensors of two different types collapse into one curve. The amplitude of the fluctuations is determined by $\N$ as a single parameter. The situation is more complex for the mean velocity. One may argue that the decay of $\langle u_x\rangle$ at $\N<170$ is still largely determined by the value of $\N$. The small values of $\langle u_x\rangle$ at $\N>170$ do not follow any clear trend. This can be related to insufficient averaging time in the experiments. The signal in figure \ref{fig:Intermitency} indicates flow dynamics that may require much longer than 600 s to measure true average values. Another possible explanation is the effect of Hartmann boundary layers, which are not uniquely identified by $\N$.

Four distinct zones characterized by different behaviours of the fluctuation amplitude and mean velocity can be identified in figure \ref{fig:N-diagram}. In zone 1, approximately at $\N<50$, the fluctuation amplitude increases and the mean velocity decreases with $\N$. The normalized standard deviation reaches the peak of $\sigma_x\approx 15U_0$ at $\N=50$. Zone 2 at $50<\N<170$ is characterized by decrease of $\sigma_x$ with $\N$ at the rate about $\sigma_x \sim \N^{-0.5}$. The mean velocity decreases with the rate higher than in zone 1. The smallest values of the fluctuation amplitude are observed at $\N=170$. In zone 3 defined approximately as $170<\N<1200$, the fluctuation amplitude grows again, initially as $\sigma_x \sim \N^{0.5}$ and then at a slower rate. We can also identify zone 4 at $\N>1200$, in which the amplitude decreases. The mean centerline velocity remains low in zones 3 and 4.

It should be stressed that the just described universal behaviour is found in the cross-section $x=1.6a$, but not in the cross-section $x=5.2a$ further downstream. We hypothesize that the effect of $\Ru$ on flow decay becomes significant at that location, so the flow state cannot be uniquely determined by $\N$ alone.

\section{Discussion and concluding remarks\label{sec:Discussion}}

In this section we attempt an explanation of the experimental results. Concluding remarks and suggestions for future work are provided at the end.

One important and evident conclusion of the experiments is that the classical picture of the transformation of a round jet in a transverse magnetic field illustrated in figures \ref{fig:JetCurandFor} and \ref{fig:JetAnnihilation} is incomplete. The theoretical arguments, on which the picture is based, fail to predict emergence of high-amplitude velocity fluctuations and modification of the mean velocity profile found at high Stuart numbers.

Another outcome of the experiment, namely that some flow properties are  accurately described  by a function of $\N$ as a single parameter is not surprising. The jet instability in the presence of a strong magnetic field is anticipated to be mainly determined by the ratio between the Lorentz and inertial forces, which is estimated by $\N$. 

The key mechanisms determining the flow evolution are the Kelvin-Helmholtz instability of the jet, transition of the flow into quasi-two-dimensional state, the Joule damping of velocity fluctuations, and the Hartmann friction. Viscous dissipation is less significant in flows with such high $\Ha$ and $\Ru$. It is ignored in the following discussion.

The jet instability is strong because of high $\Ru$ and additional perturbations introduced near the inlet by the Lorentz force in the form of transverse velocity components associated with the transformation into anisotropic form (in that way, the instability can be enhanced rather than suppressed by the magnetic field). Presence of high-amplitude fluctuations at the distance of 44 mm, i.e. about seven jet diameters from the inlet is, therefore, not surprising. 

The Joule dissipation of velocity fluctuations is strong if the fluctuations are significantly three-dimensional, but vanishes when the flow approaches quasi-two-dimensional form. The damping effect due to friction in Hartmann layers appears in quasi-two-dimensional flows, but it is, generally, weaker. These arguments have been demonstrated earlier to be a valid explanation of the so-called residual fluctuations \citep[see, e.g.,][]{kolesnikovTsinob1972,sukoriansky1986experimental,Zikanov:2019} -- high-amplitude low-frequency oscillations of velocity that appear in shear flows at high values of $\N$. One may take a more general view and consider this phenomenon as a manifestation of the general effect, found prominently in flows with thermal convection \citep{zikanov2021mixed}, of emergence of high-amplitude unsteady structures when conventional turbulence is suppressed by the magnetic field and the flow becomes quasi-two-dimensional. 

\begin{figure}
\centering
\includegraphics[width=0.99\textwidth]{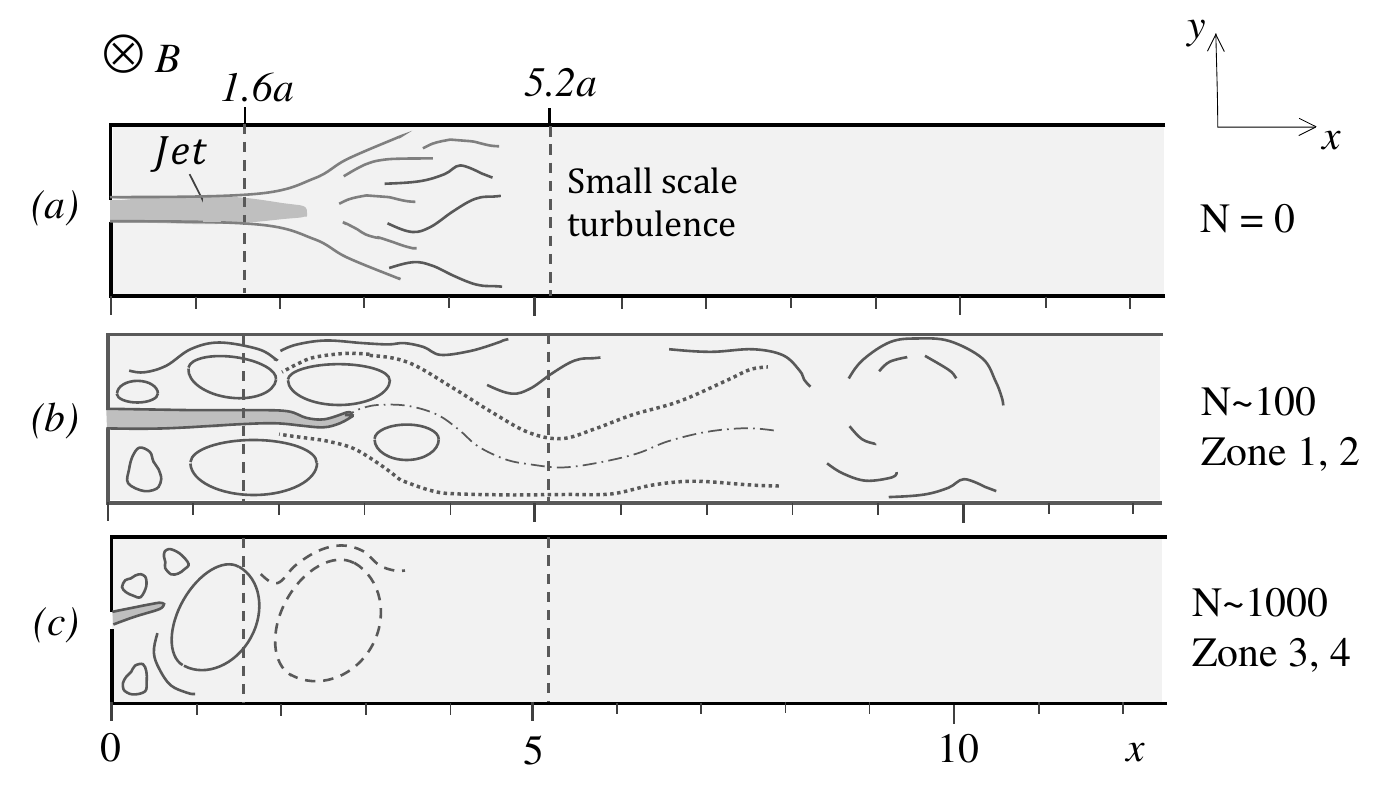}
\vskip-2mm
\caption{Sketch of the typical flow structures at various values of Stuart number. }
\label{fig:dns_re1K_vis}
\end{figure}

We will now attempt to reconstruct, based on incomplete data, the flow transformation represented by the results shown in figures \ref{fig:FirstVelocityProfile}--\ref{fig:N-diagram}. A sketch illustrating the discussion is provided in figure \ref{fig:dns_re1K_vis}.

In zones 1 and 2, approximately at $\N<170$, the mean flow  has a central jet (see figure \ref{fig:dns_re1K_vis}b). The centerline velocity decreases with  increasing $\N$. The typical amplitude of centerline velocity fluctuations increases with $\N$ in zone 1 and decreases with $\N$ in zone 2. The flow is anisotropic, but still not quasi-two-dimensional (see figure \ref{fig:FirstVelocityProfile}a).  We hypothesize that the dominant effect of the magnetic field in zone 1 is the effect consistent with the both observed trends, namely destabilization of the jet by the Lorentz force. Similarly consistent dominant effect in zone 2 is the suppression of three-dimensional velocity fluctuations by the Joule dissipation. In other words, zone 1 corresponds to the situation when the unstable three-dimensional perturbations are still growing in the measurement cross-section. Zone 2 corresponds to the situation when the magnetic field is stronger, and the growth has already been replaced by dissipation-caused decay by the time fluid arrived at the measurement site.

The sharp minimum of the fluctuation amplitude at the boundary $\N=170$ between zones 2 and 3 indicates that the flow undergoes a radical transformation. We hypothesize that zones 3 and 4 correspond to quasi-two-dimensional flow states in the measurement cross-section. The central jet is destroyed and replaced by a quasi-two-dimensional macrovortex (see figure \ref{fig:dns_re1K_vis}c). The mean velocity at the centerline is nearly zero. The amplitude of velocity fluctuations increases with $\N$ in zone 3 and decreases with $\N$ in zone 4. It must be stressed that the nature of velocity fluctuations is completely different from the nature of fluctuations in zones 1 and 2. As illustrated in figure \ref{fig:Re800Waveforms_and_spectra}, they are characterized by a steep slope of the energy power spectrum and dominance of low-frequency oscillations. The only plausible explanation of emergence of such fluctuations at high $\N$ is that they are caused by large-scale unsteady quasi-two dimensional structures associated with shear-flow instability of a macrovortex. The increase of the fluctuation amplitude with $\N$ in zone 3 can be explained by a combination of two effects promoting growth of such structures: stronger quasi-two-dimensionality and the shift of the jet-to-vortex transition upstream. The decrease of the fluctuation amplitude in zone 4 is attributed to the effect of growing friction in the Hartmann boundary layers.

The experimental results are presented in this paper for a broad range of parameters. This allows us to analyze the effect of the magnetic field and identify distinct modes of flow's behavior. At the same time, the experimental data are inevitably limited in the sense that they do not produce a complete picture of the flow.  Many aspects of the explanation proposed in this section must, therefore, be viewed as hypothetical until a confirmation in the form of numerical simulation data is available. Such simulations are a necessary and logical next step of the analysis. Their goals are (i) to produce detailed descriptions of various states of the flow and (ii) to provide data for a quantitative analysis based on detailed comparison between the experimental and numerical data.

\textbf{Funding.} Financial support was provided by the award  20-69-46067 from the Russian Science Foundation.

\textbf{Declaration of interests.} The authors report no conflict of interest.

\bibliographystyle{unsrtnat}
\bibliography{draft}

\end{document}